\documentclass[a4paper,12pt]{article}

\usepackage[latin1]{inputenc}
\usepackage[english
]{babel}
\usepackage[T1]{fontenc}
\usepackage{hyperref}

\usepackage{dsfont}
\usepackage{amsfonts}
\usepackage{amsmath}
\usepackage{amssymb}
\usepackage{amsthm}
\usepackage{eucal}
\usepackage{mathrsfs}
\usepackage[all]{xy}
\usepackage{units}

\usepackage{fancyhdr}

\usepackage[dvips]{graphicx}
\usepackage{color}

\newtheorem{teo}{\textsc{Theorem}}[section]
\newtheorem{defi}[teo]{\textsc{Definition}}
\newtheorem{propos}[teo]{\textsc{Proposition}}
\newtheorem{corol}[teo]{\textsc{Corollary}}

\newtheorem*{teo*}{\textsc{Theorem}}
\newtheorem*{defi*}{\textsc{Definition}}
\newtheorem*{corol*}{\textsc{Corollary}}

\theoremstyle{definition}
\newtheorem{ex}[teo]{\textsc{Example}}
\newtheorem*{ex*}{\textsc{Example}}
\newtheorem{rk}[teo]{\textsc{Remark}}
\newtheorem*{rk*}{\textsc{Remark}}
\newtheorem*{qst*}{\textsc{Question}}

\newcommand{\Proof}{\begin{proof}[\textsc{\bf{Proof}}]}
\newcommand{\CVD}{\end{proof}}

\newcommand{\R}{\mathbb R}
\newcommand{\N}{\mathbb N}
\newcommand{\C}{\mathbb C}                           
\newcommand{\Q}{\mathbb Q}
\newcommand{\Z}{\mathbb Z}
\newcommand{\T}{\mathbb T}

\newcommand{\sss}[1]{\CMcal{#1}}
\newcommand{\ssss}[1]{\mathcal{#1}}                  
\newcommand{\bbb}[1]{\mathscr{#1}}
\newcommand{\rrr}[1]{\mathfrak{#1}}
\newcommand{\num}[1]{\mathds {#1}}

\newcommand{\expo}[1]{\mbox{e}^{#1}}                   
\newcommand{\der}[1]{\frac{\partial}{\partial #1}}

\newcommand{\ncint}{\mathrel{{\ooalign{$\int$\cr\kern+.07em\raise.15ex\hbox{$\pmb{\scriptstyle-}$}\cr}}}}           
\newcommand{\ncpartial}{\mathrel{{\ooalign{$\partial$\cr\kern+.29em\raise.79ex\hbox{$\pmb{\scriptstyle-}$}\cr}}}}

\newcommand{\virg}[1]{\lq\lq#1\rq\rq}                
\newcommand{\ie}{{\sl i.\,e. }}
\newcommand{\eg}{{\sl e.\,g.  }}

\newcommand{\UB}{\sss{U}_\text{BF}}                  

\newcommand{\Hi}{\sss{H}}

\newcommand{\triple}{physical frame}

\DeclareMathOperator{\Ran}{Ran}

\newcommand{\I}{I}                                       

\title{ \vspace{-15mm}\textbf{\Huge{The topological Bloch-Floquet  transform  \\[3 mm] and some applications }}}
\author{\Large{$\text{G. De Nittis}^*$ and $\text{G. Panati}^{**}$
}\\
\normalsize{$^\ast$ SISSA Scuola Internazionale Superiore di Studi Avanzati, Trieste, Italy}\\
\footnotesize{\texttt{denittis@sissa.it}}\vspace{0mm}
\\
\normalsize{$^{**}$ Dipartimento di Matematica, Universit\`{a} di Roma \virg{La Sapienza}, Roma, Italy} \\
\footnotesize{\texttt{panati@mat.uniroma1.it}}\\
}

\vspace{-12mm}
\date{September 21$^{\mathrm{st}}$, 2011}
\vspace{-14mm}

\begin{document}

\maketitle              
\thispagestyle{empty}          


\begin{abstract}
\small{We investigate the relation between the symmetries of a
Schr\"{o}dinger operator and the related topological quantum
numbers. We show that, under suitable assumptions on the symmetry
algebra,  a generalization of the Bloch-Floquet transform  induces a
direct integral decomposition of the algebra of observables. More
relevantly, we prove that the generalized transform selects uniquely
the set of \virg{continuous sections} in the direct integral
decomposition, thus yielding a Hilbert bundle. The proof is
constructive and provides an explicit description of the fibers. The
emerging geometric structure is a rigorous framework for a
subsequent analysis of some topological invariants of the operator,
to be developed elsewhere \cite{DFP2}. Two running examples provide
an Ariadne's thread through the paper. For the sake of completeness,
we begin by reviewing two related classical theorems by von Neumann
and Maurin.}
\end{abstract}

\vspace{0mm}



\vspace{-5mm}

\section{Introduction}\label{sec_int}

In view of the competition between different space-scales, magnetic
Schr\"{o}dinger operators with a periodic background exhibit
striking features, as fractal spectrum and anomalous density of
states. Beyond the spectrum and the density of states, other
properties of these operators attracted the interest of physicists
and, more recently, mathematicians: the so-called \emph{Topological
Quantum Numbers} (TQN), related to observable effects whose origin
is geometric.

The prototypical example is the \emph{Hall conductance} of a
$2$-dimensional gas of non-interacting electrons in a periodic
background potential and a uniform orthogonal magnetic field. The
dynamics of the single-electron wavefunction $\psi \in L^2(\R^2,
dx\,dy)$ is governed by the Hamiltonian operator
\begin{equation}\label{Hamiltonian BL}
    H_{\beta} = \frac{1}{2}
    \left[ \left(-i \der{x} - \frac{\beta}{2} y\right)^2 +
    \left(-i \der{y} + \frac{\beta}{2} x \right)^2
    \right] + V_{\Gamma}(x,y)
\end{equation}
where $V_{\Gamma}: \R^2 \to \R$ is periodic with respect to the
lattice $\Gamma \cong \Z^2$ and $\beta$ is proportional to the
modulus of a uniform magnetic field in the direction orthogonal to
the plane. More precisely, the magnetic flux through the unit cell
of $\Gamma$ is divided by the fundamental unit of flux to obtain the
dimensionless parameter $\beta$.

If a current of intensity $I$ is forced in the $x$-direction the
charge carriers experience the Lorentz force, resulting in motion of
the carriers and a non-zero equilibrium voltage $V_{H}$ along the
$y$-direction. The \emph{Hall conductance} $\sigma_H$ is
experimentally defined as $\sigma_H = V_{H}/I$, its value depending
on both the magnetic flux $\beta$ and the density of carriers, which
depends on the Fermi energy $\mu$. While at room temperature the
measured values of $\sigma_{H}$ are in accordance with the
predictions of classical electrodynamics, the same measurement
performed at zero temperature show striking quantum features
\cite{v-klitz}, whose discovery deserved the Nobel Prize. The value
of $\sigma_{H}$, when varying either $\beta$ or $\mu$, exhibit
extremely accurate \virg{plateaux} (which are flat up to
 an accurancy of one part over $10^7$) corresponding to integer
 multiples of the constant $\nicefrac{e^2}{h}$ ($= \nicefrac{1}{2\pi}$
 in the natural units used in (\ref{Hamiltonian BL})), where $e$ is
 the charge of the electron and $h$ the Planck constant.

\begin{figure}[htbp]
\begin{center}
\includegraphics[height=7cm]{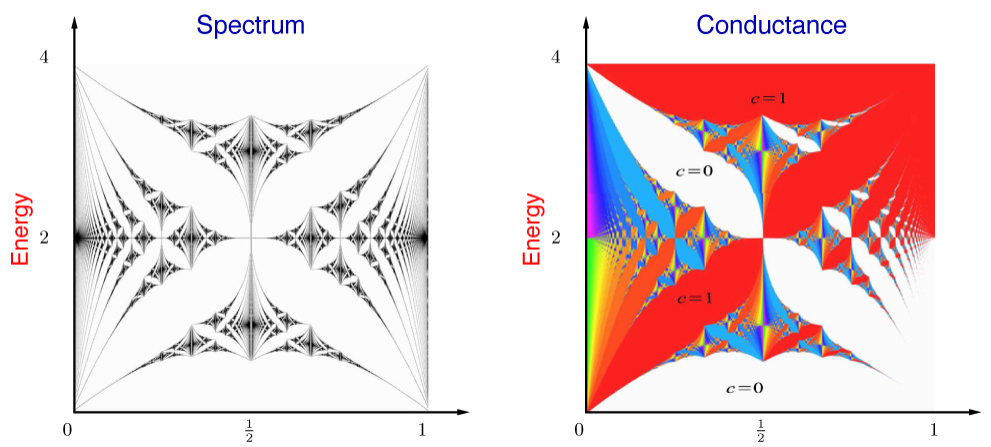}
\end{center}
\caption{{\footnotesize The black and white Hofstadter butterfly,
showing the spectrum of the Hofstadter operator as a function of the
parameter $\beta$, \emph{versus} the colored Hofstadter butterfly,
labeling the points of the resolvent set with a color corresponding
to a suitable integer (Topological Quantum Number). The colored
butterfly was originally obtained in \cite{osad}}}\label{fig1}
\end{figure}

By replacing (\ref{Hamiltonian BL}) with a simplified Hamiltonian
$K_{\beta}$ (the Hofstadter operator \cite{DP10}) a numerical
simulation becomes feasible. For $\beta \in [0,1]$ and $\mu \in
[0,4]$ the integer corresponding to $\sigma_{H}(\beta, \mu)$ is
coded by a color (warm colors for positive integers, cold colors for
negative integers and white for zero) yielding a beautiful picture
known as the \virg{colored Hofstadter butterfly} \cite{osad,avron1}.
It is assumed that $\mu$ is not in the spectrum of $K_{\beta}$,
which appears for reader's convenience on the left-hand side of the
Figure \ref{fig1}.

How are these integers (colors) related to the proprierties of the
corresponding Schr\"odinger operators? How can one read these
Topological Quantum Numbers from the Hamiltonian? The goal of this
contribution is to provide a tool to investigate the simplest
framework, namely the case of TQNs related to an abelian algebra of
symmetries $\rrr{S}$ generated by a finite family of unitary
operators. This framework includes the case of the Hall conductance
for the Hamiltonian (\ref{Hamiltonian BL}) which, as we shall see,
is related to the algebra generated by the magnetic translations. It
also includes the relevant examples of  Hofstadter and Harper
operators, which are discussed in \cite{DFP2}. In particular, the
latter paper exploits the methods here developed to prove properties
of the Hall conductance corresponding to the operator $K_{\beta}$.

We firstly recall the standard construction. To simplify the
notation, we assume that the lattice $\Gamma$ in the definition of
$H_\beta$ is simply $\Z^2$. The unitary operators $T_{\beta,1}$ and
$T_{\beta,2}$, acting on $\psi\in L^2(\R^2,dx\ dy)$ by
\begin{equation}\label{eq:magn_tras}
(T_{\beta,1}\psi)(x,y)=\expo{-i\frac{\beta}{2}y}\
\psi(x-1,y)\qquad\qquad
(T_{\beta,2}\psi)(x,y)=\expo{i\frac{\beta}{2}x}\ \psi(x,y-1),
\end{equation}
describe symmetries of the Hamiltonian \eqref{Hamiltonian BL} in the
sense that
$$
[T_{\beta,1};\, H_\beta] = 0 = [T_{\beta,2};\, H_\beta].
$$
The operators \eqref{eq:magn_tras} are known as \emph{magnetic
translations}. Unfortunately, in general $T_{\beta,1}$ and
$T_{\beta,2}$  do not commute and thus do not correspond to
simultaneously implementable symmetries (except for $\beta\in
2\pi\Z$), indeed
$$
T_{\beta,1}\ T_{\beta,2}=\expo{-i\beta}\ T_{\beta,2}\ T_{\beta,1}.
$$
In particular, the map $(n_1,n_2)\mapsto T_{\beta,1}^{n_1}\
T_{\beta,2}^{n_2}$, with $(n_1,n_2)\in\Z^2$, provides only a
projective unitary representation of the group $\Z^2$. Nevertheless,
under the so called \emph{rational flux condition} $\beta\in 2\pi\Q$
it is still possible to recover a $\Z^2$-symmetry for $H_\beta$. We
are  mainly interested in this case. Let $\beta/2\pi=p/q$ with
$p\in\Z$ and $q\in\N$ coprime. By replacing the standard lattice
$\Gamma=\Z^2$ with the \emph{super-lattice} $\Gamma':= (q\Z)^2$, one
obtains a unitary representation $(qn_1,qn_2)\mapsto
{T'_{\beta,1}}^{n_1}\ {T'_{\beta,2}}^{n_2}$  of $\Gamma'$ given by
the commuting pair of unitary operators
${T'}_{\beta,j}:=T^q_{\beta,j}$, $j=1,2$. Thus, assuming the
rational flux condition, one can still define a $\Z^2$-symmetry for
$H_\beta$ implemented by the \virg{gauged} translations
 $\{{T'}_{\beta,1},{T'}_{\beta,2}\}$.
One defines the magnetic \emph{Bloch-Floquet} (BF) transform,
initially for $\psi \in \mathcal{S}(\R^2)$, by posing
\begin{equation}\label{BF transform}
    (\UB \psi)(k, \cdot):=\sum_{n \in \Z^2} e^{- i
k \cdot n }  \big( {T'_{\beta,1}}^{n_1}\ {T'_{\beta,2}}^{n_2} \psi
\big)(\cdot), \qquad k\in\R^{d},
\end{equation}
where $n:=(n_1,n_2)$. Definition (\ref{BF transform}) extends to a
unitary operator
\begin{equation}\label{BF fibration}
\UB: L^2(\R^2) \longrightarrow \int_{\T^2}^{\oplus} \Hi(k) \, dk
\end{equation}
where $\T^2 := \R^2/\Gamma^*$ corresponds to the first (magnetic)
Brillouin zone in the physics literature, and
$$
\Hi(k) := \big\{ \varphi \in L^2_{\rm loc}(\R^2):
{T'_{\beta,1}}^{n_1}\ {T'_{\beta,2}}^{n_2}\varphi= e^{i k \cdot n}
\varphi \quad \forall n \in \Z^2 \big\}.
$$
While we focused on the bidimensional case  in view of its relevance
for the Hall conductance, the definition of the magnetic
translations and the magnetic BF transform effortlessly extend to
any dimension $d \in \N$.

In the magnetic BF representation, the Fermi projector $P_{\mu}=
\chi_{(- \infty, \mu)}(H_{\beta})$, with $\chi_I$ the characteristic
function of the set $I$,  is a decomposable operator, in the sense
that $$ \UB \, P_{\mu} \, {\UB}^{-1} = \int_{\T^2}^{\oplus}
P_{\mu}(k) \, dk.
$$ If $\mu$ lies in a spectral gap, the dimension of the range of $P_{\mu}(k)$ is
constant.  Thus it would be tempting to consider the
\emph{measurable} collection of vector spaces $\{\Ran
P_{\mu}(k)\}_{k \in \T^2}$ as a vector bundle $\sss{E}$ over $\T^2$,
and to consider its first Chern number $C_1(\sss{E}) \in \Z$ as a
topological quantum number (analogously, for $d \geq 3$ one
considers the first and the higher Chern numbers).  However, as
already emphasized, the decomposition \eqref{BF fibration} is a
measure-theoretic object, yielding only a measurable collection of
vector spaces, thus the Chern number might be undefined; even if one
circumvents this obstacle, its value might not be invariant under
unitary equivalence. In this paper we develop a construction that
yields a \emph{topological} decomposition analogous to \eqref{BF
fibration}. Moreover, the vector bundle $\sss{E}_{\rrr{S}} \to \T^d$
defined by this procedure is essentially unique, in the sense that
it is invariant under any unitary equivalence commuting with the
elements of $\rrr{S}$.

We formulate the result in a general framework: $ \sss{H}$ is a
separable Hilbert space which corresponds to the physical states;
$\rrr{A}\subset\bbb{B}(\sss{H})$ is a $C^\ast$-algebra of bounded
operators which contains the relevant physical models (the
self-adjoint elements of $\rrr{A}$ can be interpreted as
Hamiltonians); the commutant $\rrr{A}'$ (the set of all the elements
in $\bbb{B}(\sss{H})$ which commute with $\rrr{A}$) can be seen as
the {set of all the physical symmetries}; any commutative unital
$C^\ast$-algebra $\rrr{S}\subset\rrr{A}'$ describes a set of
simultaneously implementable physical symmetries.

\begin{defi}[Physical frame]
A {\upshape \triple} is a
 triple $\{\sss{H},\rrr{A},\rrr{S}\}$  where $ \sss{H}$ is a
separable Hilbert space,  $\rrr{A}\subset\bbb{B}(\sss{H})$ is a
$C^\ast$-algebra and $\rrr{S}\subset\rrr{A}'$ is a commutative
unital $C^\ast$-algebra.  The \triple\ $\{\sss{H},\rrr{A},\rrr{S}\}$
is called {\upshape irreducible} if $\rrr{S}$ is {\upshape maximal
commutative}. Two \triple s $\{\sss{H}_1,\rrr{A}_1,\rrr{S}_1\}$ and
$\{\sss{H}_2, \rrr{A}_2, \rrr{S}_2\}$ are \emph{equivalent} if there
exists a unitary map $ U:\sss{H}_1\to\sss{H}_2$ such that
$\rrr{A}_2= U \, \rrr{A}_1 \, U^{-1}$ and  \, $\rrr{S}_2=U \,
\rrr{S}_1 \, U^{-1}$.
\end{defi}

We focus on triples $\{\sss{H},\rrr{A},\rrr{S}\}$ whose
$C^\ast$-algebra $\rrr{S}$ describes symmetries with an intrinsic
group structure. In particular, in this contribution, we focus on
the case of the group $\Z^N$.

\begin{defi}[$\Z^N$-algebra]\label{Galgebra}
Let  $\Z^N \ni n \mapsto U(n) \in\bbb{U}(\sss{H})$ be a unitary
representation of  $\Z^N$  in the group $\bbb{U}(\sss{H})$ of the
unitary operators on  $\sss{H}$. The representation is
\emph{faithful} if $U(n)=\num{1}$ implies $n=0$  and is
\emph{algebraically compatible} if the  operators $\{U(n)\ :\ n
\in\Z^N \}$ are linearly
 independent in $\bbb{B}(\sss{H})$.
Let $\rrr{S}(\Z^N)$ be the unital $C^\ast$-algebra generated by
$\{U(n)\ :\ n\in \Z^N\}$.  When the representation is faithful and
algebraically compatible we  say that $\rrr{S}(\Z^N)$ is a {\upshape
$\Z^N$-algebra} in $\sss{H}$.
\end{defi}

In a nutshell, our main result is the following. Let
$\{\sss{H},\rrr{A},\rrr{S}\}$ be a physical frame with $\rrr{S}$ a
$\Z^N$-algebra satisfying the wandering property (see Definition
\ref{Def wandering}). Then there exist (and one can explicitly
construct):
\begin{itemize}
    \item a Hermitian vector bundle $\sss{E}_{\rrr{S}} \to \T^N$, whose rank is equal to
    the cardinality of the wandering system;
    \item a unitary operator $\sss{F}_{\rrr{S}}: \sss{H} \to \Gamma_{L^2}(\sss{E}_{\rrr{S}} \to
    \T^N)$, the latter being the Hilbert space consisting of the $L^2$-sections of the
    Hermitian vector
    bundle $\sss{E}_{\rrr{S}} \to \T^N$;
\end{itemize}
such that the $\ast$-subalgebra $\rrr{A}^0 \subset \rrr{A}$,
consisting of some adjointable operators in $\rrr{A}$ (see
Proposition \ref{propnew1}), satisfies
$$
\sss{F}_{\rrr{S}}\,\, \rrr{A}^0 \,\, \sss{F}_{\rrr{S}}^{-1} \subset
\Gamma(\mathrm{End}(\sss{E}_{\rrr{S}}) \to \T^N)
$$
where $\Gamma(\cdot)$ is the space of the \emph{continuous} sections
of the vector bundle appearing as the argument.  Moreover, if
$\{\sss{H},\rrr{A}_1,\rrr{S}_1\}$ and
$\{\sss{H},\rrr{A}_2,\rrr{S}_2\}$ are equivalent physical frames,
then $\sss{E}_{\rrr{S}_1}$ and $\sss{E}_{\rrr{S}_2}$ are isomorphic
Hermitian vector bundles. In particular, their topological
invariants are the same, and can thus be considered a fingerprint of
the physical frame.

\medskip

Last but not least, it is physically interesting to consider the
case of perturbed abelian symmetries, \eg to take into account the
effects of disorder and impurities in the physical system, or, in
the case of the Hamiltonian (\ref{Hamiltonian BL}), an irrational
value of the parameter $\beta/2\pi$.  In this case the algebra of
symmetries is replaced by a non-abelian $C^{\ast}$-algebra, and the
related TQNs should be investigated either with the tools of Non
Commutative Geometry \cite{bellissard}, or with other abstract
methods \cite{gruber}. On the other hand, the aim of this paper is
not to pursue the greatest generality, but rather to provide an
explicit and manageable tool to define and to compute the TQNs
hidden in the symmetries of  Sch\"odinger operators.

\goodbreak
\medskip

The paper is organized as follows:

{\bf - Section 2} introduces two relevant examples which will
reappear later, as an Ariadne's thread through the paper.

-  in {\bf Section \ref{xxy} and \ref{maurinteor}} we review two
cornerstones in the classical literature concerning the measurable
decomposition of a physical frame: the \emph{von Neumann's complete
spectral theorem} (Theorem \ref{compspec}) and the \emph{Maurin's
nuclear spectral theorem} (Theorem \ref{nucspec}).

{\bf - Section \ref{secorth}} concerns the notion of \emph{wandering
property} for a commutative $C^\ast$-algebra generated by a finite
family of  operators. This notion is of particular relevance when
the generators of the $C^\ast$-algebra are a finite set of unitary
operators. In this case we prove that the wandering property assures
that the $C^\ast$-algebra is a $\Z^N$-algebra with $N$ the number of
generators. Moreover the Gel'fand spectrum of the $C^\ast$-algebra
is forced to be the dual group of $\num{Z}^N$, i.e. the
$N$-dimensional torus $\num{T}^N$.

- in {\bf Section \ref{blochfloquetformula}}  we extend the
decomposition \eqref{BF transform} to the case of a $\Z^N$-algebra
which satisfies the wandering property. This \emph{generalized
Bloch-Floquet transform} provides a concrete recipe to decompose the
Hilbert space and the algebra $\rrr{A}$ according to the von Neumann
and Maurin theorems.

{\bf - Section \ref{emergentgeometryy}} contains our novel results:
we show that a  \emph{topological} decomposition of the algebra
$\rrr{A}$ emerges in a canonical way from the generalized
Bloch-Floquet transform, and we prove two decomposition theorems
(Theorems \ref{emerging_geo} and \ref{teonewnew}). The topological
structure is essentially unique, so the emerging information is a
fingerprint of the given \triple.

\medskip

{\textbf{Acknowledgements:} It is a pleasure to thank Gianfausto
Dell'Antonio for many stimulating discussions, and for his constant
advise and encouragement. Financial support by the {{INdAM-GNFM}
project \emph{Giovane ricercatore 2009}} is gratefully acknowledged.


\parskip 1.5mm         

\section{Some guiding examples}

We elucidate the theory with two explicit examples, while other
relevant applications are covered elsewhere \cite{DFP2}
\cite{DLandi}.

\begin{ex}[\emph{Periodic systems, part one}]\label{exper}
Let $H_{\Gamma}$ be a \emph{$\Gamma$-periodic} operator defined on
$L^2(\R^d)$. With the word $\Gamma$-periodic we mean that there
exists a linear basis $\{\gamma_1, \ldots , \gamma_d\}$ of $\R^d$
which spans the lattice $\Gamma\simeq\Z^d$ and $d$ \emph{gauged}
translations $\{T_1, \ldots, T_d \}$ defined by
$(T_j\psi)(x)=g_j(x)\ \psi(x-\gamma_j)$, where
$g_j(\cdot-\gamma)=g_j(\cdot)$ for all $\gamma\in\Gamma$ , such that
$[H_{\Gamma};T_j]=0$ for all $j=1,\ldots,d$. From the definition it
follows that $[T_i; \, T_j]=0$ for any $i,j$. Both the cases of the
magnetic translations with rational flux condition (cf. Section
\ref{sec_int}) and the \virg{genuine} translations (i.e. $g_j\equiv
1$) fit in this scheme.

 The Gel'fand-Na\v{\i}mark Theorem shows that there exists an
isomorphism between the commutative $C^\ast$-algebra
$C_0(\sigma(H_\Gamma))$ and a commutative non-unital
$C^\ast$-algebra $\rrr{A}_0(H_\Gamma)$ of bounded operators in
$\sss{H}$. The elements of $\rrr{A}_0(H_\Gamma)$ are the operators
$f(H_\Gamma)\in\bbb{B}(\sss{H})$, for  $f\in C_0(\sigma(H_\Gamma))$,
obtained via the spectral theorem. Let $\rrr{A}(H_\Gamma)$ be the
multiplier algebra of $\rrr{A}_0(H_\Gamma)$ in $\bbb{B}(\sss{H})$.
This is a unital commutative $C^\ast$-algebra which contains
$\rrr{A}_0(H_\Gamma)$ (as an essential ideal), its Gel'fand spectrum
is a (Stone-$\check{\text{C}}$ech) compactification of
$\sigma(H_\Gamma)$ and the Gel'fand isomorphism maps
$\rrr{A}(H_\Gamma)$ into the unital $C^\ast$-algebra of the
continuous and bounded functions on $\sigma(H_\Gamma)$ denoted by
$C_\text{b}(\sigma(H_\Gamma))$ (see Appendix \ref{secgelf} for
details). We assume that $\rrr{A}(H_\Gamma)$ is the $C^\ast$-algebra
of physical models.

The $C^\ast$-algebra  $\rrr{S}_T$ generated by the gauged
translations $T_j$ is clearly commutative. Since $[H_\Gamma; \,
T_j]=0$ it follows that $\rrr{S}_T\subset \rrr{A}(H_\Gamma)'$. Thus
$\{L^2(\R^d),\rrr{A}(H_\Gamma),\rrr{S}_T\}$  is a \triple. It is a
convenient model to study the properties of an electron in a
periodic medium. \hfill $\blacktriangleleft\vartriangleright$
\end{ex}

\begin{ex}[\emph{Mathieu-like Hamiltonians, part one}]\label{exmat}
Let $\T:=\R/(2\pi\Z)$ be the one-dimensional torus. In the Hilbert
space $L^2(\T)$ consider the Fourier orthonormal basis
$\{e_n\}_{n\in\Z}$ defined by
$e_n(\theta):=(2\pi)^{-\frac{1}{2}}\expo{in\theta}$. Let $\rrr{u}$
and $\rrr{v}$ be the unitary operators defined, for $g\in L^2(\T)$,
by
\begin{equation}\label{KH2}
(\rrr{u} g)(\theta):=\expo{i\theta}\ g(\theta),\ \ \ \ \ \ \ \ \ \
(\rrr{v} g)(\theta):= g(\theta-2\pi\alpha),\ \ \ \ \ \ \ \ \
\rrr{u}\rrr{v}=\expo{i2\pi\alpha}\ \rrr{v} \rrr{u}
\end{equation}
with $\alpha\in\R$. The last equation in \eqref{KH2} shows that the
unitaries $\rrr{u}$ and
 $\rrr{v}$ satisfy the  commutation relation of a \emph{noncommutative torus} with deformation parameter $\alpha$ (see \cite{boca} Chapter 1 or \cite{bon} Chapter 12 for more details).
We denote by $\rrr{A}^\alpha_\text{M}\subset\bbb{B}(L^2(\T))$ the
unital $C^\ast$-algebra generated by $\rrr{u}$, $\rrr{v}$. We call
$\rrr{A}^\alpha_\text{M}$ the \emph{Mathieu
$C^\ast$-algebra}\footnote{Such an algebra is a  representation of
the rotation $C^*$-algebra and in particular it is a faithful
representations when $\alpha\notin\Q$ \cite{boca}. Since in this
paper we focus on properties which \emph{do} depend on the
representation, we will adopt different names for images of the same
abstract $C^*$-algebra under unitarily inequivalent
representations.}
 and we refer to its elements  as Mathieu-like operators. This name is due to the fact that the
Hamiltonian $\rrr{h}:=\rrr{u}+\rrr{u}^\dag+\rrr{v}+\rrr{v}^\dag\in
\rrr{A}^\alpha_\text{M}$ appears in the well know
\emph{(almost-)Mathieu eigenvalue equation}
\begin{equation}
(\rrr{h}g)(\theta) \equiv
g(\theta-2\pi\alpha)+g(\theta+2\pi\alpha)+2\cos(\theta)g(\theta)=\varepsilon
g(\theta).
\end{equation}
The action of $\rrr{u}$ and $\rrr{v}$ on the Fourier basis is given
explicitly by $\rrr{u} e_n=e_{n+1}$ and $\rrr{v} e_n=\expo{-i2\pi
n\alpha} e_n$ for all $n\in\Z$.

We focus now on the commutant ${\rrr{A}^\alpha_\text{M}}'$ of the
Mathieu $C^\ast$-algebra. Let $\rrr{s}\in \bbb{B}(L^2(\T))$ be a
bounded operator such that
$[\rrr{s};\rrr{u}]=0=[\rrr{s};\rrr{v}_\alpha]$ and let $\rrr{s}
e_n=\sum_{m\in\Z}s_{n,m}\ e_m$ be the action of $\rrr{s}$ on the
basis vectors. The relation $[\rrr{s};\rrr{u}]=0$ implies
$s_{n+1,m+1}=s_{n,m}$ and the relation $[\rrr{s};\rrr{v}_\alpha]=0$
implies $\expo{-i2\pi(m-n)\alpha}s_{n,m}=s_{n,m}$ for all
$n,m\in\Z$.
 If $\alpha\notin\Q$ then $\expo{-i2\pi(m-n)\alpha}\neq 1$ unless $n=m$,
 hence $s_{n,m}=0$ if $n\neq m$ and the condition $s_{n+1,n+1}=s_{n,n}$
 implies that $\rrr{s}=s\num{1}$ with $s\in\C$. This shows that in the
 irrational case $\alpha\notin\Q$ the commutant of the Mathieu $C^\ast$-algebra is trivial.

To have a non trivial commutant we need to assume that
$\alpha:=\nicefrac{p}{q}$ with $p,q$ non zero integers such that
$\text{{\upshape gcd}}(q,p)=1$. In  this case the condition
$\rrr{s}\in {\big( \rrr{A}^{\nicefrac{p}{q}}_\text{M}\big)}'$
implies that $s_{n,m}\neq 0$ if and only if $m-n=k q$ for some
$k\in\Z$, moreover $s_{n,n+kq}=s_{0,kq}=:s'_k$ for all $n\in\Z$. Let
$\rrr{w}$ be the unitary operator defined on the orthonormal basis
by $\rrr{w} e_n:=e_{n+q}$, namely $\rrr{w}=(\rrr{u})^q$. The
relations for the commutant imply that $\rrr{s}\in {\big(
\rrr{A}^{\nicefrac{p}{q}}_\text{M} \big)}'$ if and only if
$\rrr{s}=\sum_{k\in\Z}s'_{k}\ {\rrr{w}}^k$. Then in the rational
case the commutant of the Mathieu $C^\ast$-algebra is the von
Neumann algebra generated in $\bbb{B}(L^2(\T))$ as the strong
closure of the family of finite polynomials in $\rrr{w}$. We will
denote by $\rrr{S}^q_\text{M}$ the unital commutative
$C^\ast$-algebra generated by $\rrr{w}$. Observe that it does not
depend on $p$. The triple
$\{L^2(\T),\rrr{A}^{\nicefrac{q}{p}}_\text{M},\rrr{S}^q_\text{M}\}$
is an example of a \triple. \hfill
$\blacktriangleleft\vartriangleright$
\end{ex}

\goodbreak

Finally, we introduce some notation which will be  useful in the
following.
\begin{rk}[\emph{Notation}]\label{Rem notation}
   The $N$-dimensional torus $\T^N:=\R^N/(2\pi\Z)^N$ is parametrized by the cube $[0,2\pi)^N$:
   for every $t=(t_1,\ldots,t_N)$ in the cube, $z(t):=(z_1(t),\ldots,z_N(t))$,
   with $z_j(t):=\expo{it_j}$, is a point of $\T^N$. The normalized Haar measure
    is $dz(t)=\nicefrac{dt_1\ldots dt_N}{(2\pi)^N}$.\hfill $\blacklozenge\lozenge$
\end{rk}


\section{The complete spectral theorem by von Neumann}\label{xxy}
The complete spectral theorem is a useful generalization of the
usual spectral decomposition of a normal operator on a Hilbert
space. It shows that the symmetries reduce the description of the full
algebra $\rrr{A}$ to a family of simpler representations.
The main tool used in the theorem is the notion of
the direct integral of Hilbert spaces (Appendix \ref{dirint}). The
\virg{spectral} content of the theorem amounts
to the characterization of the base space for the decomposition (the
\virg{set of labels}) and of the measure which glues together the
spaces so that the Hilbert space structure is preserved. This
information emerges essentially from the Gel'fand theory (Appendix
\ref{secgelf}). The definitions of decomposable and continuously
diagonal operator are reviewed in Appendix \ref{dirint}.

\begin{teo}[von Neumann's complete spectral theorem]\label{compspec}
Let  $\{\sss{H}, \rrr{A}, \rrr{S} \}$ be a physical frame and $\mu$
the basic measure carried by the spectrum $X$ of $\rrr{S}$ (see
Appendix A). Then there exist
\begin{enumerate}
\item[{\upshape a)}] a direct integral $\rrr{H}:=\int^\oplus_X\sss{H}(x)\ d\mu(x)$ with
$\sss{H}(x)\neq\{0\}$ for all $x\in X$,
\item[{\upshape b)}] a unitary map $\sss{F}_\rrr{S}:\sss{H}\to\rrr{H}$,
called {\upshape $\rrr{S}$-Fourier transform}\footnote{ According to
the terminology used in \cite{mau}.},
\end{enumerate}
such that:
\begin{enumerate}
\item[{\upshape (i)}] the unitary map $\sss{F}_\rrr{S}$ intertwines the Gel'fand isomorphism
$C(X)\ni f\stackrel{\bbb{G}}{\longmapsto} A_f\in\rrr{S}$ and the
canonical isomorphism of $C(X)$ onto the  continuously diagonal
operators $C(\rrr{H})$, i.e. the following diagram commutes
$$
\xymatrix{
&f\in C(X)\ar@{->}[dl]_{\bbb{G}}\ar@{->}[dr]&\\
 \rrr{S}\ni A_f\ar@{->}[rr]_{\sss{F}_\rrr{S}\ldots{\sss{F}_\rrr{S}}^{-1}}& & M_f(\cdot)\in C(\rrr{H})\\
}$$
\item[{\upshape (ii)}] the unitary conjugation $\sss{F}_\rrr{S} \ldots \sss{F}_\rrr{S}^{-1}$
maps the elements of \, $\rrr{A}$ to decomposable operators on
$\rrr{H}$;  more precisely, there is a family $\{ \pi_x \}_{x \in
X}$ such that $\pi_x$ is a representation of $\rrr{A}$ on
$\sss{H}(x)$ and, for every $A \in \rrr{A}$, the map $x \mapsto
\pi_x(A)$ is measurable and
$$\sss{F}_\rrr{S}\, A \,
{\sss{F}_\rrr{S}}^{-1}=\int^\oplus_X \pi_x(A)\ d\mu(x);$$
\item[{\upshape (iii)}] the representations $\pi_x$ are irreducible if and only if the physical frame $\{\sss{H}, \rrr{A}, \rrr{S} \}$ is irreducible.
\end{enumerate}
\end{teo}

\begin{rk}
For a complete proof of the above theorem one can see \cite{mau}
(Theorem 25 in Chapter I and Theorem 2 in Chapter V) or \cite{dix2}
(Theorem 1 in Part II, Chapter 6). For our purposes it is
interesting to recall how the fiber Hilbert spaces $\sss{H}(x)$ are
constructed. For $\psi,\varphi\in\sss{H}$ let
$\mu_{\psi,\varphi}=h_{\psi,\varphi}\ \mu$ be the relation between
the spectral measure $\mu_{\psi,\varphi}$ with the basic measure
$\mu$. For $\mu$-almost every $x\in X$ the value of the
Radon-Nikodym derivative $h_{\psi,\varphi}$ in $x$ defines a
semi-definite sesquilinear form on $\sss{H}$, i.e.
$(\psi;\varphi)_x:=h_{\psi,\varphi}(x)$. Let
$\sss{I}_x:=\{\psi\in\sss{H}\ :\ h_{\psi,\psi}(x)=0\}$. Then the
quotient space $\sss{H}/\sss{I}_x$ is a pre-Hilbert space and
$\sss{H}'(x)$ is defined to be the its completion. By construction
$\sss{H}'(x)\neq\{0\}$ for $\mu$-almost every $x\in X$. Let
$N\subset X$ be the $\mu$- negligible set on which $\sss{H}'(x)$ is
trivial or not well defined. Then $\rrr{H}:=\int^\oplus_X\sss{H}(x)\
d\mu(x)$ with $\sss{H}(x):=\sss{H}'(x)$ if $x\in X\setminus N$ and
$\sss{H}(x):=H$ if $x\in N$ where $H$ is an arbitrary non trivial
Hilbert space.\hfill $\blacklozenge\lozenge$
\end{rk}

Given the triple $\{\sss{H},\rrr{A},\rrr{S}\}$, the direct integral
decomposition invoked in the statement of Theorem \ref{compspec} is
essentially unique in measure-theoretic sense. The space $X$ is
unique up to homeomorphism: it agrees with the spectrum of
$C(\rrr{H})$ in such a way that the canonical isomorphism of $C(X)$
onto $C(\rrr{H})$ may be identified with the Gel'fand isomorphism.
As for the uniqueness of the direct integral decomposition, the
following result holds true  (see \cite{dix2} Theorem 3 in Part II
Chapter 6).

\begin{teo}[Uniqueness]\label{uniq}
With the notation of Theorem \ref{compspec},  let ${\nu}$ be a
positive measure  with support $X$, $\prod_{x\in X}{\sss{K}}(x)$ a
field of non-zero Hilbert spaces over $X$ endowed with a measurable
structure, ${\rrr{K}}:=\int_X^\oplus{\sss{K}}(x)\ d{\nu}(x)$,
$C({\rrr{K}})$ the commutative unital $C^\ast$-algebra of
continuously diagonal operators on ${\rrr{K}}$ and $C(X)\to
C({\rrr{K}})$ the canonical isomorphism. Let $\sss{W}$ be a unitary
(antiunitary) map from $\sss{H}$ onto ${\rrr{K}}$ transforming by
conjugation $A_f\in\rrr{S}$ into $M'_f(\cdot )\in C({\rrr{K}})$ for
all $f\in C(X)$,  in such a way that the diagram on the right hand
side commutes.
$$
\xymatrix{
&M_f(\cdot)\in C(\rrr{H})\ar@{<-}[dl]_{\sss{F}_\rrr{S}\ldots{\sss{F}_\rrr{S}}^{-1}\ }\ar@{<-}[dr]&\\
 \rrr{S}\ni A_f\ar@{->}[dr]_{\sss{W}\ldots{\sss{W}}^{-1}\ \ }\ar@{<-}[rr]^{\bbb{G}}&
 & f\in C(X)\ar@{->}[dl],\\
&M'_f(\cdot)\in C(\rrr{K})&
}\ \ \ \ \ \ \ \ \ \xymatrix{
&\rrr{H}\ar@{<-}[dl]_{\sss{F}_\rrr{S}}\ar@{-->}[dd]^{W(\cdot)}\\
 \sss{H}\ar@{->}[dr]_{\sss{W}}& \\
&\rrr{K} }$$ Then, $\mu$ and ${\nu}$ are equivalent measures (so one
can assume that $\mu={\nu}$ up to a rescaling isomorphism). Moreover
there exists a decomposable unitary (antiunitary) $W(\cdot )$ from
$\rrr{H}$ onto ${\rrr{K}}$ such that
$W(x):\sss{H}(x)\to{\sss{K}}(x)$ is a unitary (antiunitary) operator
$\mu$-almost everywhere and $\sss{W} = W(\cdot)
\circ{\sss{F}_\rrr{S}} $, i.e. the  diagram on the left hand side
commutes.
\end{teo}

\begin{corol}[Unitary equivalent triples]\label{appcor1} Let
$\{\sss{H}_1,\rrr{A}_1,\rrr{S}_1\}$ and $\{\sss{H}_2, \rrr{A}_2,
\rrr{S}_2\}$ be two equivalent \triple s and $U$ the  unitary map
which intertwines them. Let $\rrr{H}_1$ and $\rrr{H}_2$ denote the
direct integral decomposition of the two triples and let
$\sss{F}_{\rrr{S}_1}$ and $\sss{F}_{\rrr{S}_2}$ be the two
$\rrr{S}$-Fourier transforms. Then $W(\cdot ):=
\sss{F}_{\rrr{S}_2}\circ U\circ{\sss{F}_{\rrr{S}_1}}^{-1}$ is a
decomposable unitary operator from $\rrr{H}_1$ to ${\rrr{H}_2}$, so
that $W(x):\sss{H}_1(x)\to {\sss{H}_2}(x)$ is a unitary map for
$\mu$-almost every $x \in X$.
\end{corol}

\section{The nuclear spectral theorem by Maurin}\label{maurinteor}
The complete spectral theorem by von Neumann  shows that any
physical frame $\{\sss{H},\rrr{A},\rrr{S}\}$  admits a
representation in which the Hilbert space is decomposed (in a
measure-theoretically unique way) into a direct integral
$\int^\oplus_X\sss{H}(x)\ d\mu(x)$, the elements of $\rrr{S}$ are
simultaneously diagonalized  and the $C^\ast$-algebra $\rrr{A}$ is
decomposed on the fibers. The contribution of Maurin is a
characterization of the fiber spaces $\sss{H}(x)$ as {common
generalized eigenspaces} for $\rrr{S}$.

A key ingredient of  Maurin's theorem is the notion of
\emph{(nuclear) Gel'fand triple}. The latter is a triple
$\{\Phi,\sss{H},\Phi^\ast\}$ with $\sss{H}$ a separable Hilbert
space, $\Phi\subset\sss{H}$ a norm-dense subspace  such that $\Phi$
has a topology  for which it is a nuclear space and the inclusion
map $\imath:\Phi{\hookrightarrow}\sss{H}$ is continuous, and
$\Phi^\ast$ is  topological dual of $\Phi$. By identifying $\sss{H}$
with its dual space $\sss{H}^\ast$ one gets an \emph{antilinear}
injection $\imath^\ast:\sss{H}{\hookrightarrow}\Phi^\ast$. Since the
duality pairing between $\Phi$ and $\Phi^\ast$ is compatible with
the scalar product on $\sss{H}$, namely $ \langle
\imath^\ast(\psi_1); \psi_2\rangle = (\psi_1;\psi_2)_\sss{H}$
whenever $\psi_1\in \sss{H}$ and $\psi_2 \in \Phi$, we write
$\langle \psi_1; \psi_2\rangle$ for $ \langle \imath^\ast(\psi_1);
\psi_2\rangle$.
\newline
If $A$ is a bounded operator on
$\sss{H}$ such that $A^{\dag}$ leaves invariant $\Phi$ and
$A^{\dag}:\Phi\to\Phi$ is continuous with respect to the nuclear
topology of $\Phi$, one defines $ \hat A: \Phi^\ast\to\Phi^\ast$  by
posing $ \langle \hat A \, \eta ; \varphi \rangle := \langle \eta;
A^{\dag} \, \varphi \rangle$ for all $\eta\in\Phi^\ast$ and
$\varphi\in\Phi$. Then $\hat A$ is continuous and is an extension of
$A$, defined on $\sss{H}$, to $\Phi^\ast$.

Assume the notation of Theorem \ref{compspec}. Let  $\{\xi_k(\cdot
)\ :\ k \in \I\}$ be a \emph{fundamental family} of  orthonormal measurable vector
fields (see Appendix \ref{dirint}) for the direct integral $\rrr{H}$ defined by the
$\rrr{S}$-Fourier transform $\sss{F}_\rrr{S}$. Any square integrable
vector field $\varphi(\cdot )$ can be written in a unique way as
$\varphi(\cdot )=\sum_{k \in \I}\widehat{\varphi}_k(\cdot )\
\xi_k(\cdot )$ where $\widehat{\varphi}_k\in L^2(X,d\mu)$ for all
$k \in \I$. Equipped with this notation, the scalar product in
$\rrr{H}$ reads
$$
\langle\varphi(\cdot );\psi(\cdot )\rangle_\rrr{H}=\int_X\sum_{k=1}^{\text{dim}\ \sss{H}(x)}\overline{\widehat{\varphi}_k(x)}\ \widehat{\psi}_k(x)\ d\mu(x).
$$
For any $\varphi\in\sss{H}$ let $\varphi(\cdot
):=\sss{F}_\rrr{S}\varphi$ be the square integrable vector field
obtained from $\varphi$ by the $\rrr{S}$-Fourier transform. Denote
by $A_f\in\rrr{S}$ the operator associated with $f\in C(X)$ through
the Gel'fand isomorphism. One checks that
\begin{equation}\label{app7}
\widehat{\left(\sss{F}_\rrr{S}A_f\varphi\right)}_k(x)=\left(\xi_k(x);f(x)\varphi(x)\right)_x=f(x)\ \widehat{\varphi}_k(x)\ \ \ \ \ k=1,2,\ldots,\text{dim}\ \sss{H}(x).
\end{equation}
Suppose that $\{\Phi,\sss{H},\Phi^\ast\}$ is a Gel'fand triple for
the space $\sss{H}$. If $\varphi\in\Phi$ then the map
$\Phi\ni\varphi\mapsto\widehat{\varphi}_k(x):=\left(\xi_k(x);\varphi(x)\right)_x\in\C$
is linear; moreover it is possible to show that it is continuous
with respect to the nuclear topology of $\Phi$, for an appropriate
choice of $\Phi$. This means that there exists
$\eta_k(x)\in\Phi^\ast$ such that
\begin{equation}\label{app8}
\langle\eta_k(x);\varphi\rangle:=\widehat{\varphi}_k(x)=\left(\xi_k(x);\varphi(x)\right)_x\
\ \ \ \ k=1,2,\ldots,\text{dim}\ \sss{H}(x).
\end{equation}
Suppose that $A_f:\Phi\to\Phi$ is continuous with respect to the
nuclear topology for every $f \in C(X)$. Then from equations
\eqref{app7} and \eqref{app8} one has that the extended operator $
\hat A_f: \Phi^\ast\to\Phi^\ast$, namely $\langle \hat A_f \, \eta
;\varphi \rangle:=\langle\eta; A_{\overline{f}} \, \varphi \rangle$
for all $\eta\in\Phi^\ast$ and $\varphi\in\Phi$, satisfies
\begin{equation}\label{app9}
\langle \hat A_f\ \eta_k(x);\varphi\rangle=\langle
\eta_k(x);A_{\overline{f}}\,\varphi\rangle=\overline{f}(x)\
\widehat{\varphi}_k(x)=\langle f(x)\eta_k(x);\varphi\rangle\ \ \ \ \
k=1,2,\ldots,\text{dim}\ \sss{H}(x)
\end{equation}
for all $\varphi\in\Phi$. Hence,
$$ \hat A_f\ \eta_k(x)=f(x)\ \eta_k(x)  \qquad \mbox{in $\Phi^\ast$.}$$
In this sense $\eta_k(x)$ is a \emph{generalized eigenvector} for
$A_f$. These claims are made precise in the following statement.

\begin{teo}[Maurin's nuclear spectral theorem]\label{nucspec}
With the notation and the assumptions of Theorem \ref{compspec},
 let
$\{\Phi,\sss{H},\Phi^\ast\}$ be a nuclear Gel'fand triple for the
space $\sss{H}$ such that $\Phi$ is $\rrr{S}$-invariant, i.e. each
$A\in\rrr{S}$ is a continuous linear map $A:\Phi\to\Phi$. Then:
\begin{enumerate}
\item[{\upshape (i)}] for all $x\in X$ the $\rrr{S}$-Fourier transform $\left.\sss{F}_\rrr{S}\right|_x:\Phi\to\sss{H}(x)$,
$\varphi\mapsto\varphi(x)\in\sss{H}(x)$ is continuous with respect to the nuclear topology for $\mu$-almost every $x\in X$;
\item[{\upshape (ii)}] there is a family of linear functionals $\{\eta_k(x)\, :\, k=1,2,\ldots,\text{{\upshape dim}}\ \sss{H}(x)\}\subset\Phi^\ast$ such that equations \eqref{app8} and \eqref{app9} hold true for $\mu$-almost all $x\in X$;
\item[{\upshape (iii)}]  with the identification $\eta_k(x)\leftrightarrow \xi_k(x)$  the Hilbert space $\sss{H}(x)$ is (isomorphic to) a vector subspace of $\Phi^\ast$;  with this identification the $\sss{F}_\rrr{S}$-Fourier transform is defined on the dense set $\Phi$ by
\begin{equation}\label{app11}
\Phi\ni
\varphi\stackrel{\left.\sss{F}_\rrr{S}\right|_x}{\longmapsto}\sum_{k=1}^{\text{{\upshape
dim}}\ \sss{H}(x)}\langle\eta_k(x);\varphi\rangle\ \eta_k(x)\in\Phi^\ast
\end{equation}
and the scalar product in $\sss{H}(x)$ is formally defined by posing
$(\eta_k(x); \eta_j(x))_{x}:=\delta_{k,j}$;
\item[{\upshape (iv)}] under the identification in {\upshape (iii)} the spaces $\sss{H}(x)$
become the {\upshape common  generalized eigen\-spaces} of the
operators in $\rrr{S}$ in the sense that if $A_f\in\rrr{S}$ then $
\hat A_f\ \eta_k(x)=f(x)\ \eta_k(x)$ for $\mu$-almost every $x\in X$
and all $k=1,2,\ldots,\text{{\upshape dim}}\ \sss{H}(x)$.
\end{enumerate}
\end{teo}
For a proof we refer to \cite{mau} (Chapter II). The identification
at point (iii) of  Theorem \ref{nucspec} depends on the choice of a
fundamental family of orthonormal measurable vector fields
$\{\xi_k(\cdot )\, :\, k \in \I\}$  for the direct integral
$\rrr{H}$, which is clearly not unique. If $\{\zeta_k(\cdot )\, :\,
k \in \I\}$ is a second fundamental family of orthonormal measurable
vector fields for $\rrr{H}$, then there exists a decomposable
unitary map $W(\cdot )$ such that $W(x)\xi_k(x)=\zeta_k(x)$ for
$\mu$-almost every $x\in X$ and every $k \in \I$. The composition
$U:={\sss{F}_\rrr{S}}^{-1}\circ W(\cdot )\circ\sss{F}_\rrr{S}$ is a
unitary isomorphism of the Hilbert space $\sss{H}$ which induces a
linear isomorphism between the Gel'fand triples
$\{\Phi,\sss{H},\Phi^\ast\}$ and $\{\Psi,\sss{H},\Psi^\ast\}$ where
$\Psi:=U\Phi$. One checks that $\Psi$ is a nuclear space in
$\sss{H}$ with respect to the topology induced from $\Phi$ by the
map $U$ (i.e. defined by the family of seminorms
$p'_\alpha:=p_\alpha\circ U^{-1}$).  $\Psi^\ast$, the topological
dual of $\Psi$, is $\hat U \Phi^\ast$, in view of the continuity of
$U^{-1}:\Psi \to \Phi$. The isomorphism of the Gel'fand triples is
compatible with the direct integral decomposition. Indeed if
 $\vartheta_k(x)\leftrightarrow \zeta_k(x)$
is the identification between the new orthonormal basis
$\{\zeta_k(x)\, :\, k=1,2,\ldots,\text{{\upshape dim}}\
\sss{H}(x)\}$ of $\sss{H}(x)$ and a  family of linear functionals
$\{\vartheta_k(x)\, :\, k=1,2,\ldots,\text{{\upshape dim}}\
\sss{H}(x)\}\subset{\Psi}^\ast$ then equation \eqref{app8} implies
that for any $\varphi\in\Psi$
\begin{equation}\label{app12}
\langle\vartheta_k(x);\varphi\rangle:=\left(\zeta_k(x);{\varphi}(x)\right)_x=\left(\xi_k(x);W(x)^{-1}{\varphi}(x)\right)_x=\langle\eta_k(x);U^{-1}\varphi\rangle=\langle
\hat U \, \eta_k(x);{\varphi}\rangle.
\end{equation}
As a consequence, we get the following result.

\begin{propos}\label{appglob}
Up to a canonical identification of isomorphic Gel'fand triples  the
realization \eqref{app11} of the fiber spaces $\sss{H}(x)$ as
 common generalized eigen\-spaces is canonical in the sense that it
does not depend on the choice of a fundamental family of orthonormal
measurable fields.
\end{propos}
From Proposition \ref{appglob} and Corollary \ref{appcor1} it
follows that:

 \begin{corol}\label{appcor3}
Up to a canonical identification of isomorphic Gel'fand triples,
the realization \eqref{app11} of the fiber spaces $\sss{H}(x)$ as
generalized common  eigen\-spaces is preserved by a unitary
transform of the triple $\{\sss{H},\rrr{A},\rrr{S}\}$.
\end{corol}

Theorem \ref{nucspec} assumes the existence of a $\rrr{S}$-invariant
nuclear space and the related  Gel'fand triple. If
$\rrr{S}$ is generated by a countable family, such a nuclear space
does exist and there is an algorithmic procedure to construct it.
\begin{teo}[Existence of the nuclear space \cite{mau}]\label{exnuc}
Let $\{A_1,A_2,\ldots\}$ be a countable family of commuting bounded
normal operators on the separable Hilbert space $\sss{H}$, generating
the commutative $C^\ast$-algebra $\rrr{S}$. Then there exists a countable
$\rrr{S}$-cyclic system $\{\psi_1,\psi_2,\ldots\}$  which generates
a nuclear space $\Phi\subset\sss{H}$ such that: {\upshape a)} $\Phi$
is dense in $\sss{H}$; {\upshape b)} the embedding
$\imath:\Phi\hookrightarrow\sss{H}$ is continuous; {\upshape c)}
the maps $A_j^m:\Phi\to\Phi$ are continuous for all $j,m\in\N$.
\end{teo}

\begin{rk}\label{rkcyc} For the proof of Theorem \ref{exnuc}  see \cite{mau} (Chapter II, Theorem 6).
 We recall that a countable (or finite) family $\{\psi_1,\psi_2,\ldots\}$ of orthonormal
 vectors in $\sss{H}$ is a \emph{$\rrr{S}$-cyclic system}  if
 the set $\{{A^\dag}^bA^a\psi_k\ :\ k \in \I,\ \ a,b\in\N^\infty_\text{fin}\}$ is
 total in $\sss{H}$, where $\N^\infty_\text{fin}$ is the space of  $\N-$valued sequences which are
 definitely zero (i.e. $a_n=0$ for any $n \in \N \setminus I$ with $|I|< + \infty$)
 and $A^a:=A_1^{a_1}A_2^{a_2}\ldots A_N^{a_N}$ for some integer $N$.

Any $C^\ast$-algebra $\rrr{S}$ (not necessarily commutative)  has
many $\rrr{S}$-cyclic systems. Indeed one can start from any
normalized vector $\psi_1\in\sss{H}$ to build the closed subspace
$\sss{H}_1$ spanned by the action of $\rrr{S}$ on $\psi_1$. If
$\sss{H}_1\neq\sss{H}$ one can choose a second normalized vector
$\psi_2$ in the orthogonal complement of $\sss{H}_1$ to build the
closed subspace $\sss{H}_2$. Since $\sss{H}$ is separable, this
procedure produces a countable (or finite) family
$\{\psi_1,\psi_2,\ldots\}$ such that
$\sss{H}=\sss{H}_1\oplus\sss{H}_2\oplus\ldots\,$. Obviously this
construction is not unique. The nuclear space $\Phi$ claimed in
Theorem \ref{exnuc} depends on the choice of a $\rrr{S}$-cyclic
system and generally  many inequivalent choices are possible.\hfill
$\blacklozenge\lozenge$
\end{rk}
\section{The wandering property}\label{secorth}
An interesting and generally unsolved problem is the construction of
the invariant subspaces of an operator or of a family of operators.
Let $\rrr{S}$ be a  $C^\ast$-algebra contained in
$\bbb{B}(\sss{H})$. If $\psi\in\sss{H}$ then the subspace $\rrr{S}[\psi]$
generated by the action of $\rrr{S}$ on the vector $\psi$ is an
invariant subspace for the $C^\ast$-algebra. The existence of a
particular decomposition of the Hilbert space into invariant subspaces
depends on the nature of the $C^\ast$-algebra. The problem is
reasonably simple to solve for the $C^\ast$-algebras which satisfy
the wandering property.

\begin{defi}[wandering property]\label{Def wandering}
Let $\rrr{S}$ be a commutative unital $C^\ast$-algebra  generated by
the countable family $\{A_1,A_2,\ldots\}$ of commuting bounded
normal operators in a separable Hilbert space $\sss{H}$. We
say that  $\rrr{S}$ has  the {\upshape wandering property} if there
exists a (at most) countable family
$\{\psi_1,\psi_2,\ldots\}\subset\sss{H}$ of orthonormal vectors
which is $\rrr{S}$-cyclic (according to Remark \ref{rkcyc}) and such
that
\begin{equation}\label{app17'}
(\psi_k;{A^\dag}^bA^a\psi_h)_\sss{H}=\|A^a\psi_k\|^2_{\sss{H}}\ \delta_{k,h}\ \delta_{a,b}\ \ \ \ \ \ \forall\  h,k \in \I,\ \ \forall\  a,b\in\N^\infty_\text{{\upshape fin}},
\end{equation}
where $A^a:=A_1^{a_1}A_2^{a_2}\ldots A_N^{a_N}$, $\delta_{k,h}$ is
the usual Kronecker delta and $\delta_{a,b}$ is the Kronecker delta
for the multiindices $a$ and $b$.
\end{defi}
Let $\sss{H}_k:=\rrr{S}[\psi_k]$ be the Hilbert subspace generated
by the action of $\rrr{S}$ on the vector $\psi_k$. If $\rrr{S}$ has
the wandering property then the Hilbert space decomposes as
$\sss{H}=\bigoplus_{k \in \I}\sss{H}_k$ and each $\sss{H}_k$ is an
$\rrr{S}$-invariant subspace. We will refer to $\sss{H}_k$ as a \emph{
wandering subspace} and to $\{\psi_1,\psi_2,\ldots\}$ as the \emph{
wandering system}. In these subspaces each operator $A_j$  acts as a unilateral  weighted shift and this justifies
the use of the adjective \virg{wandering} (see \cite{nagy} Chapter
1, Sections 1 and 2). The wandering property implies many
interesting consequences.

\begin{propos}\label{prosistem}
Let $\rrr{S}$ be a commutative unital $C^\ast$-algebra  generated by
the (at most) countable family $\{A_1,A_2,\ldots\}$ of commuting
bounded normal operators on a separable Hilbert
space $\sss{H}$. Suppose that $\rrr{S}$ has the wandering property
with respect to the family of vectors $\{\psi_1,\psi_2,\ldots\}$,
then:
\begin{enumerate}
\item[{\upshape (i)}]  the generators are not selfadjoint, and $A_j^n \neq \num{1}$ for every $n \in \N \setminus \{0\}$;
\item[{\upshape (ii)}]  every generator which is unitary has no eigenvectors;
\item[{\upshape (iii)}]  if $\rrr{S}$ is generated by $N$ unitary operators then $\rrr{S}$ is
a $\Z^N$-algebra.
\end{enumerate}
\end{propos}
\Proof To prove (i) observe that  the  condition $A_j=A_j^\dag$
implies that $A_j\psi_k=0$ for all $\psi_k$ in the system and the
$\rrr{S}$-cyclicity imposes $A_j=0$. As for the second claim, by
setting $b=0$ and $h=k$ in  equation \eqref{app17'} one sees that
$A^a = \num{1}$ implies $a =0$.

To prove (ii) observe that if $\{U,A_1,A_2,\ldots\}$ is a set of
commuting generators for $\rrr{S}$ with $U$ unitary, then each
vector $\varphi\in\sss{H}$ can be written as
$\varphi=\sum_{n\in\Z}U^n\chi_n$ where $\chi_n= \sum_{k \in \I, \, a
\in \Z^N} \alpha_{k,a} \, A^{a}\psi_k$. Clearly
$U\varphi=\sum_{n\in\Z}U^n\chi_{n-1}$ and equation \eqref{app17'}
implies that
$\|\varphi\|^2_\sss{H}=\sum_{n\in\Z}\|\chi_n\|^2_\sss{H}$. If
$U\varphi=\lambda\varphi$, with $\lambda\in\num{S}^1$,  then a
comparison between the components provides
$\chi_{n-1}=\lambda\chi_n$, i.e. $\chi_n=\lambda^{-n}\chi_0$ for all
$n\in\Z$. This contradicts the convergence of the series expressing
the norm of $\varphi$.

To prove (iii) observe that the map $\Z^N\ni a:=(a_1,\ldots,a_N)\mapsto U^a=U_1^{a_1}\ldots U_N^{a_N}\in\bbb{U}(\sss{H})$ is a unitary representation of $\Z^N$ on $\sss{H}$. To  show
that the representation is algebraically compatible, suppose that
$\sum_{a\in\Z^N}\alpha_aU^a=0$; then from equation \eqref{app17'} it
follows that
$0=(U^b\psi_k;\sum_{a\in\Z^N}\alpha_aU^a\psi_k)_\sss{H}=\alpha_b$
for all $b\in\Z^N$, and this concludes the proof. \CVD

Proposition \ref{prosistem} shows that the  wandering property
forces a commutative $C^\ast$-algebra generated by a finite number
of unitary operators to be a $\Z^N$-algebra. This is exactly  what
happens in the cases in which we are mostly interested.

\begin{ex}[\emph{Periodic systems, part two}]\label{exwand1}
The commutative unital $C^\ast$-algebra $\rrr{S}_T$ defined in
Example \ref{exper} is generated by a  unitary faithful
representation of
 $\Z^d$ on  $L^2(\R^d)$, given by $\Z^d\ni
m \mapsto T^m\in\bbb{U}(L^2(\R^d))$ where $m:=(m_1,\ldots,m_d)$ and
$T^m:=T_1^{m_1}\ldots T_d^{m_d}$. It is easy to show that the
$C^\ast$-algebra $\rrr{S}_T$ has the wandering property. Indeed let
$\sss{Q}_0:=\{x=\sum_{j=1}^dx_j\gamma_j\ :\
-\nicefrac{1}{2}\leqslant x_j\leqslant\nicefrac{1}{2},\ \
j=1,\ldots,d \}$ be the \emph{fundamental unit cell} of the lattice
$\Gamma$ and $\sss{Q}_m:=\sss{Q}_0+m$ its translation by the lattice
vector $m:=\sum_{j=1}^dm_j\gamma_j$. Let $\{\psi_k\}_{k\in\N}\subset
L^2(\R^d)$ be a family of functions with support in $\sss{Q}_0$
providing an orthonormal basis of $L^2(\sss{Q}_0)$ up to the natural
inclusion $L^2(\sss{Q}_0)\hookrightarrow L^2(\R^d)$. This system is
$\rrr{S}_T$-cyclic since
$L^2(\R^d)=\bigoplus_{m\in\Z^d}L^2(\sss{Q}_m)$.
 Moreover, it is  wandering under the action of $\rrr{S}_T$
 since the intersection $\sss{Q}_0\cap \sss{Q}_m$ has zero measure for every  $m\neq 0$.
 The cardinality of the wandering system is $\aleph_0$.
Proposition \ref{prosistem} assures that $\rrr{S}_T$ is a
$\Z^d$-algebra. Moreover, as a consequence of Proposition
\ref{propbasicunit}, the Gel'fand spectrum of $\rrr{S}_T$  is
homeomorphic to the $d$-dimensional torus $\T^d$   and the
normalized basic measure
 is  the Haar measure $dz$ on $\T^d$.\hfill $\blacktriangleleft\vartriangleright$
\end{ex}

\begin{ex}[\emph{Mathieu-like Hamiltonians, part two}]\label{exwand2}
The unital commutative $C^\ast$-algebra
$\rrr{S}^q_\text{M}\subset\bbb{B}(L^2(\T))$  defined in Example
\ref{exmat}  is generated by a unitary faithful representation of
the group $\Z$ on the  Hilbert space $L^2(\T)$. Indeed, the map
$\Z\ni k \mapsto \rrr{w}^k\in \bbb{U}(L^2(\T))$ is an injective
group homomorphism. The set of vectors
$\{e_0,\ldots,e_{q-1}\}\subset L^2(\T)$ shows that the
$C^\ast$-algebra $\rrr{S}^q_\text{M}$  has the wandering property.
In this case the cardinality of the wandering system is $q$.
Proposition \ref{prosistem} assures that $\rrr{S}^q_\text{M}$ is a
$\Z$-algebra. Moreover, Proposition \ref{propbasicunit} will show
that  the Gel'fand spectrum of $\rrr{S}^q_\text{M}$ is homeomorphic
to the $1$-dimensional torus $\T$ and  the normalized basic measure
on the spectrum coincides with the Haar measure $dz$ on $\T$. The
first claim agrees with the fact that the Gel'fand spectrum of
$\rrr{S}^q_\text{M}$ coincides with the (Hilbert space) spectrum of
$\rrr{w}$, the generator of the $C^\ast$-algebra, and
$\sigma(\rrr{w})=\T$. The claim about the basic measure agrees with
the fact
 that the vector $e_0$ is cyclic  for the
 commutant of $\rrr{S}^q_\text{M}$ (which is the von Neumann algebra generated by ${\rrr{A}^{\nicefrac{p}{q}}_\text{M}}$).
 Indeed, a general result (see Appendix \ref{secgelf}) assures
 that the spectral measure $\mu_{e_0,e_0}$ provides the basic measure. To determine $\mu_{e_0,e_0}$ let
 $\ssss{P}(\rrr{w}):=\sum_{k\in\Z}\alpha_k\rrr{w}^k$ be any element of $\rrr{S}^q_\text{M}$. From the definition of spectral measure it follows that
\begin{equation}\label{eq1}
\alpha_0=(e_0;\ssss{P}(\rrr{w})e_0)=\int_\T\ssss{P}(z)\
d\mu_{e_0,e_0}(z)=\sum_{k\in\Z}\alpha_k\int_0^{2\pi}\expo{ik t}\
d\widetilde{\mu}_{e_0,e_0}(t).
\end{equation}
where the measure $\widetilde{\mu}_{e_0,e_0}$ is related to
${\mu}_{e_0,e_0}$ by the change of variables $\T\ni z\mapsto
t\in[0,2\pi)$ defined by $z:=\expo{it}$ according to Remark \ref{Rem
notation}. Equation \eqref{eq1} implies that
$\widetilde{\mu}_{e_0,e_0}$  agrees with $\nicefrac{dt}{2\pi}$ on
$C(\T)$, namely  the basic measure
 $\mu_{e_0,e_0}$ is the normalized Haar
measure.\hfill$\blacktriangleleft\vartriangleright$
\end{ex}

In the relevant cases of commutative unital $C^\ast$-algebras
generated by a finite set of unitary operators the wandering
property provides a useful  characterization of  the Gel'fand
spectrum and the basic measure. We firstly introduce some notation
and terminology. Let $\num{G}$ be a discrete group and
$\ell^1(\num{G})$ be the set of sequences $c=\{c_g\}_{g\in\num{G}}$
such that $\|c\|_{\ell^1}=\sum_{g\in\num{G}}|c_g|<+\infty$. Equipped
with the convolution product $(c\ast
d)_g:=\sum_{h\in\num{G}}c_{h}d_{g-h}$ and involution
$c^\dag:=\{\overline{c}_{-g}\}_{g\in\num{G}}$ , $\ell^1(\num{G})$
becomes a unital Banach $\ast$-algebra called the \emph{group
algebra} $\num{G}$. The latter is not a $C^\ast$-algebra since the
norm $\|\cdot\|_{\ell^1}$ does not verify the $C^\ast$-condition
$\|c \ast c^\ast\|_{\ell^1}=\|c\|^2_{\ell^1}$. In general there
exist several inequivalent ways to complete $\ell^1(\num{G})$ to a
$C^\ast$-algebra by introducing suitable $C^\ast$-norms. Two of
these $C^\ast$-extensions are of particular interest. The first is
obtained as the completion of $\ell^1(\num{G})$ with respect to the
\emph{universal enveloping norm}
$$
\|c\|_u:=\sup\{\|\pi(c)\|_{\bbb{\sss{H}}}\ :\ \pi:\ell^1(\num{G})\to\bbb{B}(\sss{H}) \ \text{is a}\ \ast-\text{representation}
\}.
$$
The resulting abstract $C^\ast$-algebra, denoted by
$C^\ast(\num{G})$, is called the \emph{group $C^\ast$-algebra} of
$\num{G}$ (or \emph{enveloping} $C^\ast$-algebra).

The second relevant extension is obtained by means of the concrete
representation of the elements $\ell^1(\num{G})$ as (convolution)
multiplicative operators on the Hilbert space  $\ell^2(\num{G})$. In
other words, for any $\xi
=\{\xi_g\}_{g\in\num{G}}\in\ell^2(\num{G})$ and
$c=\{c_g\}_{g\in\num{G}}\in\ell^1(\num{G})$ one defines the
representation $\pi_r:\ell^1(\num{G})\to\bbb{B}(\ell^2(\num{G}))$ as
$$
\pi_r(c)\xi:=c\ast\xi=\left\{\sum_{h\in\num{G}}c_{h}\xi_{g-h}\right\}_{g\in\num{G}}.
$$
The representation $\pi_r$, known as \emph{left regular
representation}, is injective. The norm
$\|c\|_r:=\|\pi_r(c)\|_{\bbb{B}(\ell^2(\num{G}))}$ defines a new
$C^\ast$-norm on $\ell^1(\num{G})$, called \emph{reduced norm}, and
a new $C^\ast$-extension denoted by $C^\ast_r(\num{G})$ and called
\emph{reduced group $C^\ast$-algebra}. Since $\|\cdot\|_r\leq
\|\cdot\|_u$ it follows that $C^\ast_r(\num{G})$ is
$\ast$-isomorphic to a quotient $C^\ast$-algebra of
$C^\ast(\num{G})$. Nevertheless, if the group $\num{G}$ is abelian,
one has the relevant isomorphism
$C^\ast_r(\num{G})=C^\ast(\num{G})\simeq C(\widehat{\num{G}})$ where
$\widehat{\num{G}}$ denotes the dual (or character) group of
$\num{G}$. For more details the reader can refer to \cite{dix1}
(Chapter 13) or \cite{davidson}(Chapter VII).

\begin{propos}\label{propbasicunit}
Let $\sss{H}$ be a separable Hilbert space and $\rrr{S} \subset
\bbb{B}(\sss{H})$ a unital commutative $C^\ast$-algebra generated by
a finite family $\{U_1,\ldots,U_N\}$ of unitary operators. Assume
the wandering property. Then:
\begin{enumerate}
\item[{\upshape (i)}]  the Gel'fand spectrum of $\rrr{S}$ is homeomorphic to the $N$-dimensional torus $\T^N$;
\item[{\upshape (ii)}]  the basic measure of $\rrr{S}$ is the normalized Haar measure $dz$ on $\T^N$.
\end{enumerate}
\end{propos}
\Proof We use the short notation $ U^a = U_1^{a_1}\ldots U_N^{a_N}$
for any $a=(a_1,\ldots,a_N)\in\Z^N$.

To prove (i) one notices that the map
$F:\ell^1(\Z^N)\to\bbb{B}(\sss{H})$, defined by
$F(c):=\sum_{a\in\Z^N}c_a\ {U}^{a}$, is a $\ast$-representation of
$\ell^1(\Z^N)$ in $\bbb{B}(\sss{H})$. As in the proof of
Proposition \ref{prosistem}, one exploits the wandering property to
see that for any $c \in \ell^1(\Z^N)$, $\sum_a c_a U^a = 0$ implies
$c=0$. Thus $F$ is a faithful representation. Moreover
$\|F(c)\|_{\bbb{B}(\sss{H})}\leqslant\|c\|_{\ell^1}$ for all
$c\in\ell^1(\num{C})$. Finally, the unital $\ast$-algebra
$\rrr{L}^1(\Z^N):=F(\ell^1(\Z^N))\subset\bbb{B}(\sss{H})$ is dense
in $\rrr{S}$ (with respect to the operator norm), since it does
contain the polynomials in $U_1, \ldots, U_N$, which are a dense
subset of $\rrr{S}$.

In view of the fact that $\Z^N$ is abelian, to prove (i) it is
sufficient to show that $\rrr{S}\simeq C^\ast_r(\Z^N)$. Since
$\ell^1(\Z^N)$ and $\rrr{L}^1(\Z^N)$ are isomorphic Banach
$\ast$-algebras, and $\rrr{L}^1(\Z^N)$ is dense in $\rrr{S}$, the
latter claims follows if one proves that
$\|c\|_r=\|F(c)\|_{\bbb{B}(\sss{H})}$ for any $c\in\ell^1(\Z^N)$.
Let $\{\psi_k\}_{k \in \I}$ be the wandering system of vectors for
$\rrr{S}$. The wandering property assures that the closed subspace
$\rrr{S}[\psi_k]=:\sss{H}_k\subset\sss{H}$ is isometrically
isomorphic to $\ell^2(\Z^N)$, with unitary isomorphism given by
$\sss{H}_k\ni\sum_{a\in\Z^N}\xi_aU^a\psi_k\mapsto\{\xi_a\}_{a\in\Z^N}\in\ell^2(\Z^N)$.
Then, due to the mutual orthogonality of the spaces $\sss{H}_k$,
there exists a unitary map
$\sss{R}:\sss{H}\to\bigoplus_{k \in \I}\ell^2(\Z^N)$ which extends all
the isomorphisms above. A simple computation shows that
$\sss{R}F(c)\sss{R}^{-1}=\bigoplus_{k \in \I}\pi_r(c)$ for any
$c\in\ell^1(\Z^N)$. Since $\sss{R}$ is isometric, it follows that
$\|F(c)\|_{\bbb{B}(\sss{H})}=\|\bigoplus_{k \in \I}\pi_r(c)\|_{\bigoplus_{k \in \I}\ell^2}=
\|\pi_r(c)\|_{\ell^2}$, which is exactly the definition of the norm
$\|c\|_r$.

To prove (ii)  let $\mu_k:=\mu_{\psi_k,\psi_k}$ be the spectral
measure defined by the vector of the wandering system $\psi_k$. The
Gel'fand isomorphism identifies the generator $U_j\in\rrr{S}$ with
$z_j\in C(\T^N)$. It follows  that for every $a \in \Z^N$ one has
\begin{equation}\label{eq2}
\delta_{a,0}=(\psi_k;U^a\psi_k)=\int_{\T^N}z^a\
d\mu_k(z):=\int_0^{2\pi}\ldots\int_{0}^{2\pi}z_1^{a_1}(t)\ldots
z_N^{a_N}(t)\ d\widetilde{\mu}_k(t),
\end{equation}
where the measure $\widetilde{\mu}_{k}$ is related to ${\mu}_{k}$ by
the change of variables $\T^N\ni z\mapsto t\in[0,2\pi)^N$ defined by
$z:=\expo{it}$ according to Remark \ref{Rem notation}. Equation
\eqref{eq2} shows that for all $k$ the spectral measure
$\widetilde{\mu}_k$ agrees with  $dz(t):=\nicefrac{dt_1\ldots
dt_N}{(2\pi)^N}$.

Let $A_f$ be the element of $\rrr{S}$ whose image via the Gel'fand
isomorphism is the function $f\in C(\T^N)$. Then
$$
(U^b\psi_j;A_fU^a\psi_k)_\sss{H}=\delta_{j,k}(\psi_k;A_fU^{a-b}\psi_k)_\sss{H}=\int_{\T^N}f(z)\
\delta_{j,k}\ z^{a-b}\ dz.
$$
So the spectral measure $\mu_{U^b\psi_j,U^a\psi_k}$ is related to
the Haar measure $dz$ by the function $\delta_{j,k}\ z^{a-b}$. Let
$\varphi:=\sum_{k \in \I, a\in\N^N}\alpha_{a,k}\ U^a\psi_k$ be any
vector in $\sss{H}$. Notice that, in view of the wandering property,
one has $\alpha_{k, a} \in \ell^2(\N) \otimes \ell^2(\Z^N)$. Then a
direct computation shows that
$\mu_{\varphi,\varphi}(z)=h_{\varphi,\varphi}(z)\ dz$, where $
h_{\varphi,\varphi}(z)=\sum_{k \in \I}|F^{(k)}_\varphi(z)|^2$ with $
F^{(k)}_\varphi(z):=\sum_{a\in\N^N}\alpha_{k,a} z^a. $ Since
$F^{(k)}_\varphi\in L^2(\T^N)$ , one has $|F^{(k)}_\varphi|^2\in
L^1(\T^N)$. Let
$h^{(M)}_{\varphi,\varphi}(z)=\sum_{k=0}^M|F^{(k)}_\varphi(z)|^2$.
Since   $h^{(M+1)}_{\varphi,\varphi}\geqslant
h^{(M)}_{\varphi,\varphi}\geqslant0$  and
$\int_{\T^N}h^{(M)}_{\varphi,\varphi}(z)\
dz=\sum_{k=0}^M\sum_{a\in\N^N}|\alpha_{k,a}|^2\leqslant\|\varphi\|^2_\sss{H}$
for all $M$, one concludes by the monotone convergence theorem that
$h_{\varphi,\varphi}\in L^1(\T^N)$.\CVD


Not every commutative $C^\ast$-algebra generated by a faithful
unitary representation of $\Z^N$  has a wandering system. In this
situation, even if the spectrum is still a torus, the basic measure
can be inequivalent to the Haar measure, as illustrated by the
following example.

\begin{ex}\label{appnongoex}
Let $R_\alpha$ be the unitary operator on $L^2(\R^2)$ which
implements a rotation around the origin of the angle $\alpha$, with
$\alpha\notin2\pi\Q$. Since $R^N_\alpha=R_{N\alpha}\neq\num{1}$ for
every integer $N$, it follows that the commutative unital
$C^\ast$-algebra $\rrr{R}_\alpha$ generated by $R_\alpha$ is a
$\Z$-algebra. The Gel'fand spectrum of $\rrr{R}_\alpha$, which
coincides with the spectrum of $R_\alpha$, is $\T$. Indeed, the
vector $\psi_N(\rho,\phi):=\expo{iN\phi}f(\rho)$ (in polar
coordinates) is an eigenvector corresponding to the eigenvalue
$\expo{iN\alpha}$. The spectrum of $R_\alpha$ is the closure of
$\{\expo{iN\alpha}\ :\ N\in\Z\}$, which is $\T$ in view of the
irrationality of $\alpha$. The existence of eigenvectors excludes
the existence of a wandering system (see Proposition
\ref{prosistem}). Moreover, since $R_\alpha$ has point spectrum it
follows that the basic measure is not the Haar measure. Indeed, the
spectral measure $\mu_{\psi_N,\psi_N}$ corresponding to the
eigenvector $\psi_N$ is the \emph{Dirac measure} concentrated in
$\{\expo{iN\alpha}\} \subset \C$. \hfill
$\blacktriangleleft\vartriangleright$
\end{ex}

\section{The generalized Bloch-Floquet transform}\label{blochfloquetformula}

The aim of this section is to provide a general algorithm for
constructing the direct integral decomposition of a commutative
$C^\ast$-algebra which appears in the von Neumann's complete
spectral theorem.
In this approach a relevant role will be played by the wandering
property. We consider a
 commutative unital $C^\ast$-algebra
$\rrr{S}$ on a separable Hilbert space $\sss{H}$ generated by the
finite family $\{U_1,U_2,\ldots,U_N\}$ of unitary operators
admitting a wandering system $\{\psi_k\}_{k \in \I}\subset\sss{H}$.
According to the results of Section \ref{secorth}, $\rrr{S}$ is a
$\Z^N$-algebra with Gel'fand spectrum  $\T^N$ and with the Haar
measure $dz$ as basic measure.


\subsection*{Construction of the wandering nuclear space}

The existence of a wandering system makes possible the explicit construction of a
$\rrr{S}$-invariant nuclear space, which we call the \emph{wandering nuclear space}.

Consider the orthonormal basis $\{U^a\psi_k\}_{k \in \I,a\in\Z^N}$,
where $\{\psi_k\}_{k \in \I}$ is the wandering system, and denote by
$\sss{L}\subset\sss{H}$ the family of all finite linear combinations
of the vectors of this basis. For every integer $m\geqslant0$ denote
by $\sss{H}_m$ the finite dimensional Hilbert space generated by the
finite set of vectors $\{U^a\psi_k \,:\, 0 \leqslant k \leqslant m,
\, 0\leqslant |a| \leqslant m\}$, where $|a|:=|a_1|+\ldots+|a_N|$.
Obviously $\sss{H}_m\subset\sss{L}$. Let $D_m$ denote   the
dimension of the space $\sss{H}_m$. If
$\varphi=\sum_{k \in \I,a\in\Z^N}\alpha_{k,a}U^a\psi_k$ is any element
of $\sss{H}$ then  the formula
\begin{equation}\label{app17}
p_m^2(\varphi):=D_m \sum_{ { \scriptsize \begin{array}{c} 0 \leqslant k \leqslant m \\  0 \leqslant |a| \leqslant m  \\ \end{array}}} |(U^a\psi_k;\varphi)_\sss{H}|^2
=D_m\sum_{{ \scriptsize \begin{array}{c} 0 \leqslant k \leqslant m \\  0 \leqslant |a| \leqslant m  \\ \end{array}}} |\alpha_{k,a}|^2,
\end{equation}
defines a seminorm  for every $m\geqslant0$. From \eqref{app17} it
follows that $p_m\leqslant p_{m+1}$ for all $m$. The countable
family of seminorms $\{p_m\}_{m\in\N}$ provides a locally convex
topology for the vector space $\sss{L}$. Let  $\Sigma$ denote the
pair $\left\{\sss{L},\{p_m\}_{m\in\N}\right\}$, i.e. the vector
space $\sss{L}$ endowed with  the locally convex topology induced by
the seminorms \eqref{app17}. $\Sigma$ is a complete and metrizable
(i.e. Fr\'echet) space. However, for our purposes, we need a
topology on $\sss{L}$ which is strictly stronger than the metrizable
topology induced by the seminorms \eqref{app17}.

The quotient space $\Phi_m:=\sss{L}/\ssss{N}_m$,
with $\ssss{N}_m:=\{\varphi\in\sss{L}\ :\
p_m(\varphi)=0\}$, is isomorphic to the finite dimensional vector
space $\sss{H}_m$, hence it is nuclear and Fr\'echet. This follows
immediately observing that the norm $p_m$ on $\Phi_m$ coincides, up
to the positive constant $\sqrt{D_m}$, with the usual Hilbert norm.
Obviously $\Phi_m\subset\Phi_{m+1}$ for all  $m\geqslant0$ and the
topology of $\Phi_m$ agrees with the topology inherited from
$\Phi_{m+1}$, indeed
$\left.p_{m+1}\right|_{\Phi_m}=\sqrt{\frac{D_{m+1}}{D_m}}\ p_m$. We
define $\Phi$ to be $\bigcup_{m\in\N}\Phi_m$ (which is
$\sss{L}$ as a set) endowed with the \emph{strict inductive limit
topology} which is the strongest topology which makes  all
 injections $\imath_m:\Phi_m\hookrightarrow\Phi$ continuous. The space
$\Phi$ is called a \emph{LF-space} (according to the definition of
\cite{treves} Chapter 13) and it is a nuclear space since it is the
strict inductive limit of nuclear spaces (see \cite{treves}
Proposition 50.1). We will say that $\Phi$ is the \emph{wandering
nuclear space} defined by the $\Z^N$-algebra $\rrr{S}$ on the
wandering system $\{\psi_k\}_{k \in \I}$.

\begin{propos}\label{propnuc}
The wandering nuclear space $\Phi$ defined by the previous construction satisfies
all the properties stated in Theorem \ref{exnuc}.
\end{propos}
\Proof
 A linear map $\jmath:\Phi\to \Psi$, with $\Psi$ is an arbitrary locally convex
 topological vector space, is continuous if and only if the restriction
 $\left.\jmath\right|_{\Phi_m}$ of $\jmath$ to $\Phi_m$ is continuous for
 each $m\geqslant0$ (see \cite{treves} Proposition 13.1).
 This implies that the canonical embedding $\imath:\Phi\hookrightarrow\sss{H}$ is continuous, since its
 restrictions are  linear operators defined on finite dimensional spaces.
 The linear maps $U^a:\Phi\to\Phi$ for all $a\in\N^N$ are also continuous for the same reason.
 Finally $\Phi$ is norm-dense in $\sss{H}$ since  as a set it is the dense domain $\sss{L}$.
\CVD

\subsection*{The transform}

We are now in  position to define the \emph{generalized
Bloch-Floquet transform} $\sss{U}_{\rrr{S}}$ for the
$C^\ast$-algebra $\rrr{S}$. The Gel'fand spectrum of $\rrr{S}$ is
$\T^N$
 and the Gel'fand isomorphism associates to the generator $U_j$ the function $z_j\in C(\T^N)$,
 according to the notation of Remark \ref{Rem notation}.
For all $t\in [0,2\pi)^N$ and for all $\varphi\in\Phi$ we define
(formally for the moment) the Bloch-Floquet transform of $\varphi$
at point $t$  as
\begin{equation}\label{app13}
\boxed{
\qquad \Phi\ni\varphi\stackrel{\sss{U}_{\rrr{S}}|_t}{\longmapsto}(\sss{U}_{\rrr{S}}\varphi)(t):=
\sum_{a\in\Z^N}^{\textcolor[rgb]{1.00,1.00,1.00}{a\in\Z^N}} z^{-a}(t)\ U^a\varphi \qquad
}
\end{equation}
 where  $z^a(t):= e^{i a_1 t_1} \ldots e^{i a_N t_N}$  and $U^a:=U_1^{a_1}\ldots U_N^{a_N}$.
 The structure of equation \eqref{app13} suggests that
  $(\sss{U}_{\rrr{S}}\varphi)(t)$ is a common generalized eigenvector for  the elements
  of $\rrr{S}$, indeed a formal computation shows that
\begin{equation}\label{app20}
U_j(\sss{U}_{\rrr{S}}\varphi)(t)=z_j(t)\sum_{
a\in\Z^N}{z^{-1}_j(t)z^{-a}(t)} U_j \, U^a\varphi=\expo{it_j}\
(\sss{U}_{\rrr{S}}\varphi)(t).
\end{equation}

\noindent This guess is clarified by the following result.

\goodbreak

\begin{teo}[Generalized Bloch-Floquet transform]\label{trasbloc}
Let $\rrr{S}$ be a $\Z^N$-algebra in the separable Hilbert space $\sss{H}$ with generators
 $\{U_1,\ldots,U_N\}$ and wandering system $\{\psi_k\}_{k \in \I}$, and let $\Phi$
 be the corresponding wandering nuclear space.
Then the {\upshape generalized Bloch-Floquet
transform} \eqref{app13} defines an injective linear map from the
nuclear space $\Phi$ into its topological dual $\Phi^\ast$ for every
$t\in[0,2\pi)^N$. More precisely, the transform
$\left.\sss{U}_{\rrr{ S}}\right|_t$ maps $\Phi$ onto a subspace
$\Phi^\ast(t)\subset\Phi^\ast$ which is a common generalized
eigenspace for the commutative $C^\ast$-algebra $\rrr{S}$, \ie
$$
\hat U_j \, (\sss{U}_{\rrr{S}}\varphi)(t) = \expo{it_j}\,
(\sss{U}_{\rrr{S}}\varphi)(t) \qquad \mbox{ in } \,\, \Phi^\ast.
$$
The map
$\left.\sss{U}_{\rrr{S}}\right|_t:\Phi\to\Phi^\ast(t)\subset\Phi^\ast$
is a continuous linear isomorphism, provided $\Phi^\ast$ is endowed
with the weak-$\ast$ topology.
\end{teo}
\Proof
 We need to verify that the right-hand side of  \eqref{app13} is well defined as a linear functional on $\Phi$. Any vector $\varphi\in\Phi$ is a finite linear combination $\varphi=\sum^\text{fin}_{k \in \I}\sum^\text{fin}_{b\in\Z^N} \alpha_{k,b}\ U^b\psi_k$ (the complex numbers $\alpha_{k,b}$ are different from zero only for a finite set  of the values of the index $k$ and the multiindex $b$).
Let $\phi=\sum_{h\in\N}^\text{fin}\sum_{c\in\N^N}^\text{fin} \beta_{h,c}\ U^{c}\psi_{h}$ be another element in $\Phi$.
The linearity of the dual pairing between $\Phi^\ast$ and $\Phi$ and the compatibility of the pairing with the Hermitian structure of $\sss{H}$ imply
\begin{equation}\label{app14}
\langle(\sss{U}_{\rrr{S}}\varphi)(t);\phi \rangle:=\sum_{k \in \I}^\text{fin}\sum_{b,c\in\Z^N}^\text{fin}\overline{\alpha}_{k,b}\beta_{k,c}\left(\sum_{a\in\Z^N}z^a(t)
\ (U^{a+b}\psi_k;U^c\psi_k)_\sss{H}\right)
\end{equation}
where in the right-hand side we used the orthogonality between the
spaces generated by $\psi_k$ and $\psi_h$ if $k\neq h$. Without
further conditions equation \eqref{app14} is a finite sum in $k,b,c$
(this is simply a consequence of the fact that $\varphi$ and $\phi$
are \virg{test functions}) but it is an infinite sum in $a$ which
generally does not converge. However, in view of the wandering
property one has that $(U^{a+b}\psi_k;U^c\psi_k)_\sss{H}=
\delta_{a+b,c}$, so that  \eqref{app14} reads
\begin{equation}\label{app15}
\langle(\sss{U}_{\rrr{S}}\varphi)(t);\phi \rangle
=\sum_{k \in \I}^\text{fin}\sum_{b,c\in\Z^N}^\text{fin}\overline{\alpha}_{k,b}\beta_{k,c}\
z^c(t)z^{-b}(t).
\end{equation}
Let $C_{\varphi;k}:=\sum_{b\in\Z^N}^\text{fin}\left|{\alpha}_{k,b}\right|$ and  $C_\varphi:=\max_{k \in \I}\{C_{\varphi;k}\}$ (which is well defined since the set contains only a finite numbers of non-zero elements). An easy computation shows that
 \begin{align*}
|\langle(\sss{U}_{\rrr{S}}\varphi)(t);\phi \rangle| &\leqslant
\sum_{k \in \I}^\text{fin}C_{\varphi,k}\left(\sum_{c\in\Z^N}^\text{fin}|\beta_{k,c}|\right)\leqslant
C_\varphi\sum_{k \in \I}^\text{fin}\sum_{c\in\Z^N}^\text{fin}|\beta_{k,c}|\,
.
\end{align*}
Let $m\geqslant0$ be the smallest integer such that $\phi\in\Phi_m$.
The number of the coefficients $\beta_{k,c}$ different from zero is
smaller than the dimension $D_m$ of $\Phi_m$. Using the
Cauchy-Schwarz inequality one has
\begin{equation}\label{app21}
|\langle(\sss{U}_{\rrr{S}}\varphi)(t);\phi \rangle|\leqslant
C_\varphi\sqrt{D_m}\left(\sum_{k \in \I}^\text{fin}\sum_{c\in\N^N}^\text{fin}|\beta_{k,c}|^2\
\right)^{\frac{1}{2}}= C_\varphi\ p_m(\phi).
\end{equation}
The inequality \eqref{app21} shows that the linear map
$(\sss{U}_{\rrr{S}}\varphi)(t):\Phi\to\C$ is continuous when it is
restricted to each finite dimensional space $\Phi_m$. Since $\Phi$
is endowed with the strict inductive limit topology, this is enough
to assure that  $(\sss{U}_{\rrr{S}}\varphi)(t)$ is a continuous
linear functional on $\Phi$. So, in view of \eqref{app21},
$(\sss{U}_{\rrr{S}}\varphi)(t)\in\Phi^\ast$ for all $t\in
[0,2\pi)^N$ and for all $\varphi\in\Phi$.

The linearity of the map
$\left.\sss{U}_{\rrr{S}}\right|_t:\Phi\to\Phi^\ast$ is immediate and
from equation \eqref{app15} it follows that
$(\sss{U}_{\rrr{S}}\varphi)(t)=0$ (as functional) implies that
$\alpha_{k,b}=0$ for all $k$ and $b$, hence $\varphi=0$. This proves
the injectivity. To prove the continuity of the map
$\left.\sss{U}_{\rrr{S}}\right|_t:\Phi\to\Phi^\ast$, in view of the
strict inductive topology on $\Phi$, we  only need to check the
continuity of the maps
$\left.\sss{U}_{\rrr{S}}\right|_t:\Phi_m\to\Phi^\ast$ for all
$m\geqslant0$. Since $\Phi_m$ is a finite dimensional vector space
with norm $p_m$, it is sufficient to prove that the norm-convergence
of the sequence $\varphi_n\to0$ in $\Phi_m$ implies the weak-$\ast$
convergence $(\sss{U}_{\rrr{S}}\varphi_n)(t)\to0$ in $\Phi^\ast$,
i.e. $|\langle(\sss{U}_{\rrr{S}}\varphi_n)(t);\phi \rangle|\to0$ for
all $\phi\in\Phi$. As inequality \eqref{app21} suggests, it is
enough to show that $C_{\varphi_n}\to0$. This is true since
$\varphi_n:=\sum_{0<k,|b|\leqslant m} \alpha^{(n)}_{k,b}\
U^{b}\psi_{k} \to0$ in $\Phi_m$ implies $\alpha^{(n)}_{k,b}\to0$.

Finally, since the map $U^{-a}=(U^a)^\dag$ is continuous on $\Phi$
for all $a\in\Z^N$ then $\hat{(U^{a})}:\Phi^\ast\to\Phi^\ast$
defines a continuous map  which extends the operator $U^a$
originally defined on $\sss{H}$. In this context the equation
\eqref{app20} is meaningful and shows that
$\Phi^\ast(t):=\left.\sss{U}_{\rrr{S}}\right|_t(\Phi)\subset\Phi^\ast$
is a space of common generalized eigenvectors for the elements of
$\rrr{S}$. \CVD
\subsection*{The decomposition}

The wandering system  $\{\psi_k\}_{k \in \I}$ generates under the
Bloch-Floquet transform a special family of elements of $\Phi^\ast$,
denoted by
\begin{equation}\label{app22}
\boxed{ \qquad
\zeta_k(t):=(\sss{U}_{\rrr{S}}\psi_k)(t)=\sum_{ a\in\Z^N}^{\textcolor[rgb]{1.00,1.00,1.00}{a\in\Z^N}}z^{-a}(t)\ U^a\psi_k\ \qquad \forall\ k \in \I.
\qquad }
\end{equation}

The injectivity of the map $\sss{U}_{\rrr{S}}$ implies that the
functionals $\{\zeta_k(t)\}_{k \in \I}$ are linearly independent for
every $t$. If
$\varphi=\sum_{k \in \I}^\text{fin}\sum_{b\in\Z^N}^\text{fin}
\alpha_{k,b}\ U^b\psi_k$ is any element in $\Phi$ then a simple
computation shows that
\begin{equation}\label{app23}
(\sss{U}_{\rrr{S}}\varphi)(t)=\sum_{k \in \I}^\text{fin}\sum_{b\in\Z^N}^\text{fin}\alpha_{k,b}\sum_{
a\in\N^N}z^{-a}(t)\
U^{a+b}\psi_k=\sum_{k \in \I}^\text{fin}f_{\varphi;k}(t)\ \zeta_k(t)
\end{equation}
where $f_{\varphi;k}(t):=\sum_{b\in\Z^N}^\text{fin} \alpha_{k,b}\
z^b(t)$. The equalities in \eqref{app23} should be interpreted in
the sense of \virg{distributions}, \ie elements of $\Phi^\ast$. The
functions $f_{\varphi;k}:\T^N\to\C$, for all $k \in \I$, are finite
linear combinations of continuous functions, hence continuous.
Equation \eqref{app23} shows that any subspace $\Phi^\ast(t)$ is
generated by finite linear combinations of the functionals
\eqref{app22}. For every $t\in [0,2\pi)^N$ we denote by $\sss{K}(t)$
the space of the elements of the form $\sum_{k \in \I}\alpha_k\
\zeta_k(t)$ with $\{\alpha_k\}_{k \in \I}\in\ell^2(\I)$. This is a
Hilbert space with the inner product induced by the isomorphism with
$\ell^2(\I)$. In other words the inner product is induced by the
\virg{formal} conditions $(\zeta_k(t); \zeta_h(t))_t:=\delta_{k,h}$.
All the Hilbert spaces $\sss{K}(t)$ have the same dimension which is
the cardinality of the system $\{\psi_k\}_{k \in \I}$.

\begin{propos}\label{equidim}
For all $t\in[0,2\pi)^N$ the inclusions
$\Phi^\ast(t)\subset\sss{K}(t)\subset\Phi^\ast$ hold true. Moreover the
generalized Bloch-Floquet transform $\left.\sss{U}_{\rrr{S}}\right|_t$ extends
to a unitary isomorphism between the Hilbert space
$\num{H}\subset\sss{H}$ spanned by the orthonormal system
$\{\psi_k\}_{k \in \I}$ and the Hilbert space $\sss{K}(t)\subset\Phi^\ast$
spanned by $\{\zeta_k(t)\}_{k \in \I}$ (assumed as orthonormal basis).
\end{propos}
\Proof The first inclusion $\Phi^\ast(t)\subset\sss{K}(t)$ follows from the definition.
For the second inclusion we need to prove that $\omega(t):=\sum_{k \in \I}\alpha_k\ \zeta_k(t)$ is
a continuous functional  if $\{\alpha_k\}_{k \in \I}\in\ell^2(\N)$. Let
$\phi=\sum_{0\leqslant h,|c|\leqslant m} \beta_{h,c}\ U^{c}\psi_{h}$ be an  element of $\Phi_m\subset\Phi$ then, from the
sesquilinearity of the dual pairing and the Cauchy-Schwarz inequality it follows that
\begin{equation}\label{eq5}
|\langle\omega(t);\phi
\rangle|^2\leqslant\left(\sum_{k \in \I}|\alpha_k|\
|\langle(\sss{U}_{\rrr{S}}\psi_k)(t);\phi
\rangle|\right)^2\leqslant\|\alpha\|_{\ell^2}^2\sum_{k \in \I}|\langle(\sss{U}_{\rrr{S}}\psi_k)(t);\phi
\rangle|^2
\end{equation}
where $\|\alpha\|^2_{\ell^2}=\sum_{k \in \I}|\alpha_k|^2<\infty$. From
equation \eqref{app14} it is clear that
$\langle(\sss{U}_{\rrr{S}}\psi_k)(t);\phi \rangle=0$ if
$\psi_k\notin\Phi_m$, then equation \eqref{app21}  and
$C_{\psi_k}=1$ imply $|\langle\omega(t);\phi
\rangle|\leqslant\|\alpha\|_{\ell^2}\sqrt{m}\ p_m(\phi)$. This
inequality shows that $\omega(t)$ is a continuous functional when it
is restricted to each subspace $\Phi_m$ and, because the strict
inductive limit topology, this  proves that $\omega(t)$ lies in
$\Phi^\ast$.

As for the second claim, consider $\omega_n(t):=\sum_{0\leqslant
k\leqslant n}\alpha_k\ \zeta_k(t)$. Obviously
$\omega_n(t)=(\sss{U}_{\rrr{S}}\varphi_n)(t)\in\Phi^\ast(t)$ since
$\varphi_n:=\sum_{0\leqslant k\leqslant n}\alpha_k\ \psi_k\in\Phi$.
Moreover the inequality \eqref{eq5} can be used to show that
$(\sss{U}_{\rrr{S}}\varphi_n)(t)\to\omega(t)$ when $n\to\infty$ with
respect to the weak-$\ast$ topology of $\Phi^\ast$. This enables us
to define $\omega(t):=(\sss{U}_{\rrr{S}}\varphi)(t)$ for all
$\varphi:=\sum_{k \in \I}\alpha_k\ \psi_k\in \num{H}$. The generalized
Bloch-Floquet transform acts as a unitary isomorphism between
$\num{H}$ and $\sss{K}(t)$ with respect to the Hilbert structure
induced in $\sss{K}(t)$ by the orthonormal basis
$\{\zeta_k(t)\}_{k \in \I}$.\CVD

\begin{teo}[Bloch-Floquet spectral decomposition]\label{BFspectral}
Let $\rrr{S}$ be a $\Z^N$-algebra in the separable Hilbert space
$\sss{H}$ with generators $\{U_1,\ldots,U_N\}$, wandering system
$\{\psi_k\}_{k \in \I}$ and wandering nuclear space $\Phi$.  The
generalized Bloch-Floquet transform $\sss{U}_{\rrr{S}}$, defined on
$\Phi$ by equation \eqref{app13}, induces a direct integral
decomposition of the Hilbert space $\sss{H}$ which is equivalent (in
the sense of Theorem \ref{uniq}) to the decomposition of  von
Neumann's theorem (Theorem \ref{compspec}). Moreover, the spaces $\sss{K}(t)$
spanned in $\Phi^\ast$ by the functionals \eqref{app22} provide an
explicit realization for the family of
 common eigenspaces of $\rrr{S}$
appearing in Maurin's theorem (Theorem \ref{nucspec}).
\end{teo}
\Proof Proposition \ref{propbasicunit} assures that the Gel'fand
spectrum of $\rrr{S}$ is the $N$-dimensional torus $\T^N$ and the
basic measure agrees with the normalized Haar measure $dz$. On the
field of Hilbert spaces $\prod_{t\in \T^N}\sss{K}(t)$ we can
introduce a measurable structure by the fundamental family of
orthonormal vector fields $\{\zeta_k(\cdot )\}_{k \in \I}$ defined
by \eqref{app22}. For all $\varphi\in\Phi$ the generalized
Bloch-Floquet transform defines a square integrable vector field
$(\sss{U}_{\rrr{S}}\varphi)(\cdot
)\in\rrr{K}:=\int_{\T^N}^\oplus\sss{K}(t)\ dz(t)$. Indeed equation
\eqref{app23} shows that
$(\sss{U}_{\rrr{S}}\varphi)(t)\in\sss{K}(t)$ for any $t$ and
$\|(\sss{U}_{\rrr{S}}\varphi)(t)\|_t^2=\sum_{k \in
\I}^\text{fin}|f_{\varphi;k}(t)|^2$ is a continuous function (finite
sum of continuous functions) hence integrable on $\T^N$. In
particular
\begin{align*}
\|(\sss{U}_{\rrr{S}}\varphi)(\cdot)\|^2_\rrr{K}&=\int_{\T^N}\|(\sss{U}_{\rrr{S}}\varphi)(t)\|_t^2\
dz(t)=\sum_{k \in
\I}\int_{\T^N}\underbrace{\left(\sum_{b,c\in\Z^N}^\text{fin}\overline{\alpha}_{k,b}\alpha_{k,c}\
z^{c-b}(t)\right)}_{=|f_{\varphi;k}(x)|^2}\
dz(t)=\|\varphi\|^2_\sss{H}.
\end{align*}
In view of the density of $\Phi$,  $\sss{U}_{\rrr{S}}$ can be
extended to an isometry from $\sss{H}$ to $\rrr{K}$.

It remains to show that $\sss{U}_{\rrr{S}}$ is surjective. Any
square integrable vector field $\varphi(\cdot)\in\rrr{K}$  is
uniquely characterized by its expansion on the basis
$\{\zeta_k(\cdot )\}_{k \in \I}$, i.e. $\varphi(\cdot)=\sum_{k\in
N}\widehat{\varphi}_k(\cdot)\ \zeta_k(\cdot )$ where
$\{\widehat{\varphi}_k(t)\}_{k \in \I}\in\ell^2(\N)$ for all
$t\in[0,2\pi)^N$. The condition
$$
\|\varphi(\cdot)\|^2_\rrr{K}=\int_{\T^N}\sum_{k \in
\I}|\widehat{\varphi}_k(t)|^2\ dz(t)<+\infty
$$ shows that $\widehat{\varphi}_k\in L^2(\T^N)$
for all $k \in \I$. Let $\widehat{\varphi}_k(t)
=\sum_{b\in\Z^N}\alpha_{k,b}z^b(t)$ be the Fourier expansion of
$\widehat{\varphi}_k$. Since
$$
\sum_{k \in \I}\sum_{b\in\Z^N}|\alpha_{k,b}|^2 = \sum_{k \in
\I}\|\widehat{\varphi}_k\|^2_{L^2(\T^N)}=\|\varphi(\cdot)\|^2_\rrr{K}<+\infty
$$
it follows that $\{\alpha_{k,b}\}_{k \in \I,b\in\Z^N}$ is an
$\ell^2$-sequences and the mapping
\begin{equation}\label{eq6}
\varphi(\cdot)=\sum_{k\in
\N}\sum_{b\in\Z^N}\alpha_{k,b}\ z^b(\cdot)\  \zeta_k(\cdot )\ \ \ \ \stackrel{{\sss{U}_{\rrr{S}}}^{-1}}{\longmapsto} \ \ \ \  \
\varphi:=\sum_{k \in \I}\sum_{b\in\Z^N}\alpha_{k,b}\ U^b\psi_k
 \end{equation}
defines an element  $\varphi\in\sss{H}$ starting from the vector
field $\varphi(\cdot)\in\rrr{K}$. It is immediate to check that
$\sss{U}_{\rrr{S}}$ maps $\varphi$ in $\varphi(\cdot)$, hence
$\sss{U}_{\rrr{S}}$ is surjective.

If $A_f\in\rrr{S}$ is an operator associated with the continuous
function $f\in C(\T^N)$ via the Gel'fand isomorphism, then
$\sss{U}_{\rrr{S}}A_f{\sss{U}_{\rrr{S}}}^{-1}\varphi(\cdot)=f(\cdot)\varphi(\cdot)$,
i.e. $\sss{U}_{\rrr{S}}$ maps $A_f\in\rrr{S}$ in $M_f(\cdot)\in
C(\rrr{K})$. This allows us to apply the Theorem \ref{uniq} which
assures that the direct integral $\rrr{K}$ coincides, up to a
decomposable unitary transform, with the spectral decomposition of
$\rrr{S}$ established in Theorem \ref{compspec}. \CVD

\goodbreak

The generalized Bloch-Floquet transform $\sss{U}_{\rrr{S}}$  can be
seen as a \virg{computable} realization of the abstract
$\rrr{S}$-Fourier transform $\sss{F}_\rrr{S}$. From Proposition
\ref{equidim} and from  general results about direct integrals (see
\cite{dix2} Part II, Chapter 1, Section 8) one obtains the following
identifications:
\begin{equation}
\sss{H}\
\xymatrix@1{\ar@{->}[rr]^(0.5){\sss{U}_{\rrr{S}}\ldots{\sss{U}_{\rrr{S}}}^{-1}}&&}\
\int_{\T^N}^\oplus\sss{K}(t)\ dz(t)\simeq\int_{\T^N}^\oplus\num{H}\
dz(t)\simeq L^2(\T^N,\num{H}).
\end{equation}
Since the dimension of $\num{H}$ is the cardinality of the wandering
system chosen to define the Bloch-Floquet transform, and since
Theorem \ref{uniq} assures  that the direct integral decomposition
is essentially unique (in measure theoretic sense), one has the
following:
\begin{corol}\label{corocard}
Any two wandering systems associated with a $\Z^N$-algebra $\rrr{S}$
have the same cardinality. Any two  wandering systems for  $\rrr{S}$
are intertwined  by  a unitary operator which commutes with
$\rrr{S}$.
\end{corol}
\begin{ex}[\emph{Periodic systems, part three}]
In the case of  Example \ref{exper}, the Bloch-Floquet transform is
the usual one (see \cite{kuc}, \cite{Pan})
$$
(\sss{U}_{\rrr{S}_T}\varphi)(t,y):=\sum_{m\in\Gamma}z^{-m}(t)\,
T^m\varphi(y)=\sum_{m\in\Gamma}\expo{-im_1t_1}\ldots\expo{-im_dt_d}\varphi(y-m),
$$
where $m:=\sum_{j=1}^dm_j\gamma_j$, for all $\varphi$ in the
wandering nuclear space $\Phi\subset L^2(\R^d)$, built according to
Proposition \ref{propnuc} from any orthonormal basis of
$L^2(\sss{Q}_0)$. The fiber spaces in the direct integral
decomposition are all unitarily equivalent to $L^2(\sss{Q}_0)$ hence
the Hilbert space decomposition is
$$
L^2(\R^d)\
\xymatrix@1{\ar@{->}[rr]^(0.5){\sss{U}_{\rrr{S}_T}\ldots{\sss{U}_{\rrr{S}_T}}^{-1}}&&}
\ \int_{\T^d}^\oplus L^2(\sss{Q}_0)\ dz(t).
$$
\hfill $\blacktriangleleft\vartriangleright$
\end{ex}
\begin{ex}[\emph{Mathieu-like Hamiltonians, part three}]\label{exnew1}
In this case  the wandering nuclear space $\Phi$ is  the set of the
finite linear combinations of the Fourier basis $\{e_n\}_{n\in\Z}$
and for all $g(\theta)=\sum_{n\in\Z}^\text{fin}\alpha_n
\expo{in\theta}$ in $\Phi$ the Bloch-Floquet transform is
$$
(\sss{U}_{\rrr{S}^q_\text{M}}\ g)(\theta,
t):=\sum_{m\in\Z}\expo{-imt}\
\rrr{w}^mg(\theta)=\sum_{n\in\Z}^\text{fin}\alpha_n\left(\sum_{m\in\Z}\expo{i[n\theta+m(q\theta-t)]}\right).
$$
The collection $\zeta_k(\cdot;t)\in\Phi^\ast$,  given by
$\zeta_k(\theta;t):=\expo{ik\theta}\sum_{m\in\Z}\expo{im(q\theta-t)}$
with $k=0,\ldots,q-1$, defines a fundamental family of orthonormal
fields. The fiber spaces in the direct integral decomposition are
all unitarily equivalent to $\C^q$ hence the Hilbert space
decomposition is
$$
L^2(\T)\
\xymatrix@1{\ar@{->}[rr]^(0.5){\sss{U}_{\rrr{S}^q_\text{M}}\ldots{\sss{U}_{\rrr{S}^q_\text{M}}}^{-1}}&&}\
\int_{\T}^\oplus \C^q\ dz(t).
$$
The images of the generators $\rrr{u}$ and $\rrr{v}$ under the map
$\sss{U}_{\rrr{S}^q_\text{M}} \ldots
\sss{U}_{_{\rrr{S}^q_\text{M}}}^{-1}$ are the two $t$ dependent
$q\times q$ matrices
$$
\rrr{u}(t):=\left(
\begin{array}{cccc}
0 &   &  & \expo{it} \\
1 &   \ddots &  &  \\
 &\ddots    & \ddots &  \\
 0 &  & 1 & 0
\end{array}\right)\ \ \ \ \ \ \ \ \ \  \ \ \rrr{v}(t):=\left(
\begin{array}{cccc}
1&   &  &  \\
 &   \expo{-i2\pi\frac{p}{q}} &  &  \\
 &    & \ddots &  \\
  &  &  & \expo{-i2\pi\frac{p}{q}(q-1)}
\end{array}\right).
$$
For every  $t\in\T$ the matrices $\rrr{u}(t)$ and $\rrr{v}(t)$
generate a faithful irreducible representation of the
$C^\ast$-algebra $\rrr{A}^{\nicefrac{p}{q}}_\text{M}$ on the Hilbert
space $\C^q$ (see \cite{boca} Theorem 1.9).\hfill
$\blacktriangleleft\vartriangleright$
\end{ex}

\goodbreak


\medskip

\section{Emergent geometry}\label{emergentgeometryy}

From a  geometric viewpoint, the field of Hilbert spaces $\rrr{F}:=\prod_{x\in X}\sss{H}(x)$
can be regarded as a pseudo vector-bundle $\bbb{E}\stackrel{\pi}{\longrightarrow}X$,
where
\begin{equation}\label{eq10}
\bbb{E}:=\bigsqcup_{x\in X}\sss{H}(x)
\end{equation}
is the disjoint union of the Hilbert spaces $\sss{H}(x)$. The use of
the prefix \virg{pseudo} refers to the fact that more ingredients
are needed to turn $\bbb{E}\stackrel{\pi}{\longrightarrow}X$ into a
vector bundle. First of all, the map $\pi$ must be continuous, which
requires a topology on $\bbb{E}$. As a first attempt, assuming that
$\sss{H}(x) \subset \Phi^\ast$ for every $x \in X$, one might
consider $\bbb{E}\stackrel{\pi}{\longrightarrow}X$  as a sub-bundle
of the trivial vector bundle
$X\times\Phi^\ast\stackrel{\pi}{\longrightarrow}X$, equipped with
the topology induced by the inclusion, so that
$\bbb{E}\stackrel{\pi}{\longrightarrow}X$ becomes a topological
bundle whose fibers are Hilbert spaces. However, nothing ensures
that the Hilbert space topology defined fiberwise is compatible with
the topology of $\bbb{E}$, a necessary condition to have a
meaningful topological theory.
\subsection*{Geometric vs. analytic viewpoint}
We begin our analysis with the definition of topological fibration of Hilbert spaces. Following
\cite{fell1} (Chapter II, Section 13) we pose the following
\begin{defi}[Geometric viewpoint: Hilbert bundle] \label{hilbbund}
A {\upshape Hilbert bundle} is the datum of a topological Hausdorff spaces $\bbb{E}$ (the {\upshape total space}),
a compact Hausdorff space $X$ (the {\upshape base space}) and a map $\bbb{E}\stackrel{\pi}{\longrightarrow}X$
(the {\upshape canonical projection}) which is a continuous open surjection such that:
\begin{enumerate}
\item[{\upshape a)}] for all $x\in X$ the fiber $\pi^{-1}(x)\subset\bbb{E}$ is a Hilbert space;
\item[{\upshape b)}] the map $\bbb{E}\ni p\longmapsto \|p\|\in\C$ is continuous;
\item[{\upshape c)}] the operation $+$ is continuous as a function on $\bbb{S}:=\{(p,s)\in\bbb{E}\times\bbb{E}\ :\ \pi(p)=\pi(s)\}$ to $\bbb{E}$;
\item[{\upshape d)}] for each $\lambda\in\C$ the map $\bbb{E}\ni p\longmapsto \lambda\ p\in\bbb{E}$ is continuous;
\item[{\upshape e)}] let $0_x$ be the null vector in the Hilbert space $\pi^{-1}(x)$; for each $x\in X$, the collection of all subsets of $\bbb{E}$ of the form $\bbb{U}(O,x,\varepsilon)
:=\{p\in\bbb{E}\ :\ \pi(p)\in O,\ \ \|p\|<\varepsilon\}$, where $O$ is a neighborhood of $x$ and $\varepsilon>0$, is a basis of neighborhoods of $0_x\in\pi^{-1}(x)$ in $\bbb{E}$.
\end{enumerate}
\end{defi}

\medskip

We denote by the short symbol $\bbb{E}_{\pi,X}$ the Hilbert
bundle $\bbb{E}\stackrel{\pi}{\longrightarrow}X$. A \emph{section}
of $\bbb{E}_{\pi,X}$ is a function  $\psi:X\to\bbb{E}$ such that
$\pi\circ \psi=\text{id}_X$. We denote by $\Gamma(\bbb{E}_{\pi,X})$
the set of all \emph{continuous sections} of $\bbb{E}_{\pi,X}$. As
shown in \cite{fell1}, from Definition \ref{hilbbund} it follows
that: (i) the scalar multiplication $\C\times\bbb{E}\ni(\lambda,
p)\mapsto\lambda\ p\in\bbb{E}$ is continuous; (ii) the open sets of
$\bbb{E}$, restricted to a fiber $\pi^{-1}(x)$, generate the Hilbert
space topology of $\pi^{-1}(x)$; (iii) the set
$\Gamma(\bbb{E}_{\pi,X})$ has the structure of a (left)
$C(X)$-module. In other words, the definition of Hilbert bundle includes all the
conditions which a \virg{formal} fibration such  as \eqref{eq10} needs to
fulfill to be a topological fibration with a topology compatible
with the Hilbert structure of the fibers. In this sense the Hilbert
bundle is the \virg{geometric object} of our interest.

However, the structure that emerges in a natural way from the Bloch-Floquet decomposition (Theorem \ref{BFspectral}) is more easily understood from the analytic viewpoint. Switching the focus from the total space $\bbb{E}$ to the space of sections $\rrr{F}$, the relevant notion is that
of \emph{continuous field of Hilbert spaces}, according to \cite{dix1} (Section 10.1) or \cite{dix-dou} (Section 1).

\begin{defi}[Analytic viewpoint: continuous field of Hilbert spaces]
Let $X$ be a compact Hausdorff space and
$\rrr{F}:=\prod_{x\in X}\sss{H}(x)$  a field of Hilbert spaces. A \emph{continuous structure on  $\rrr{F}$} is the datum of a linear subspace $\Gamma\subset \rrr{F}$ such that:
\begin{enumerate}
\item[{\upshape a)}] for each $x\in X$ the set $\{\sigma(x)\ :\ \sigma(\cdot)\in\Gamma\}$ is dense in $\sss{H}(x)$;
\item[{\upshape b)}] for any $\sigma(\cdot)\in\Gamma$ the map $X\ni x\mapsto \|\sigma(x)\|_x\in\R$ is continuous;
\item[{\upshape c)}] if $\psi(\cdot)\in\rrr{F}$ and if for each $\varepsilon>0$ and each $x_0\in X$, there is some $\sigma(\cdot)\in\Gamma$ such that  $\|\sigma(x)-\psi(x)\|_x<\varepsilon$ on a neighborhood of $x_0$, then $\psi(\cdot)\in\Gamma$.
\end{enumerate}
\end{defi}
We  denote by the short symbol $\rrr{F}_{\Gamma,X}$ the field of
Hilbert spaces $\rrr{F}$ endowed with the continuous structure
$\Gamma$. The elements of $\Gamma$ are called  \emph{continuous
vector fields}. The condition b) may be replaced by the requirement
that for any $\sigma(\cdot),\varrho(\cdot)\in\Gamma$, the function
$X\ni x\mapsto (\sigma(x);\varrho(x))_x\in\C$ is continuous.
Condition c) is called \emph{locally uniform closure} and guarantees
that the linear space $\Gamma$ is stable under multiplication by continuous functions on $X$. This
condition implies that $\Gamma$ is a \emph{(left) $C(X)$-module}. A
\emph{total set} of continuous vector fields for
$\rrr{F}_{\Gamma,X}$ is a subset $\Lambda\subset\Gamma$ such that
$\Lambda(x):=\{\sigma(x)\ :\ \sigma(\cdot)\in\Lambda\}$ is dense in
$\sss{H}(x)$ for all $x\in X$. The continuous field of Hilbert
spaces is said to be \emph{separable} if it has a countable total
set of continuous vector fields.

The link between the notion of continuous field of Hilbert spaces and that of Hilbert bundle is clarified by the following result.

\begin{propos}[Equivalence between geometric and analytic viewpoint \cite{dix-dou} \cite{fell1}]\label{vecbund}
Let $\rrr{F}_{\Gamma,X}$ be a continuous field of Hilbert spaces
over the compact Hausdorff space $X$. Let
$\bbb{E}(\rrr{F}_{\Gamma,X}):=\bigsqcup_{x\in X}\sss{H}(x)$ be the
disjoint union of the Hilbert spaces $\sss{H}(x)$ and $\pi$ the
canonical surjection of $\bbb{E}(\rrr{F}_{\Gamma,X})$ onto $X$. Then
there exists a unique topology $\ssss{T}$ on
$\bbb{E}(\rrr{F}_{\Gamma,X})$ making
$\bbb{E}(\rrr{F}_{\Gamma,X})\stackrel{\pi}{\longrightarrow}X$ a
Hilbert bundle over $X$ such that all the continuous vector fields
in $\rrr{F}_{\Gamma,X}$ are continuous sections of
$\bbb{E}(\rrr{F}_{\Gamma,X})$. Moreover, every Hilbert bundle comes
from a continuous field of Hilbert spaces.
\end{propos}

For the proof we refer to \cite{dix-dou} (Section 2) or \cite{fell1}
(Chapter II, Theorem 13.18). We say that the set
$\bbb{E}(\rrr{F})$ endowed with the topology $\ssss{T}$ and the
canonical surjection $\pi$ is the Hilbert bundle {associated} with
the continuous structure $\Gamma$ of $\rrr{F}$.

\subsection*{Triviality, local triviality and vector bundle structure}

A Hilbert bundle is a generalization of a (infinite dimensional) vector bundle,
in the sense that some other extra conditions are needed in order to turn it into a
 genuine vector bundle. For the axioms
of vector bundle we refer to \cite{lang2}. The most relevant missing condition is the \emph{local triviality property}.

Two Hilbert bundles $\bbb{E}_{\pi,X}$ and $\bbb{F}_{\tau,X}$ over the same base space $X$ are said
 to be { (isometrically) isomorphic} if there exists a homeomorphism
 $\varTheta:\bbb{E}\to\bbb{F}$ such that a) $\tau\circ \varTheta=\pi$,
 b) $\varTheta_x:=\left.\varTheta\right|_{\pi^{-1}(x)}$ is a unitary map from
 the Hilbert space $\pi^{-1}(x)$ to the Hilbert space ${\tau}^{-1}(x)$.
 From the definition it follows that if the Hilbert bundles $\bbb{E}_{\pi,X}$ and $\bbb{F}_{\tau,X}$
 are isomorphic then the map $\Gamma(\bbb{E}_{\pi,X})\ni\sigma\mapsto\varTheta\circ\sigma\in\Gamma(\bbb{F}_{\tau,X})$
is one to one. A Hilbert bundle is said to be \emph{trivial} if it
is isomorphic to the \emph{constant} Hilbert bundle
$X\times\num{H}\to X$ where $\num{H}$ is a fixed Hilbert space. It
is called \emph{locally trivial} if for every $x\in X$ there is a
neighborhood $O$ of $x$ such that the \emph{reduced Hilbert bundle}
$\left.\bbb{E}\right|_O:=\{p\in \bbb{E}:\, \pi(p) \in O
\}=\pi^{-1}(O)$ is isomorphic to the constant Hilbert bundle
$O\times\num{H}\to O$. Two continuous fields of Hilbert spaces
$\rrr{F}_{\Gamma,X}$ and $\rrr{G}_{\Delta,X}$ over the same space
$X$ are said to be (isometrically) isomorphic if the associated
Hilbert bundles $\bbb{E}(\rrr{F}_{\Gamma,X})$ and
$\bbb{E}(\rrr{G}_{\Delta,X})$ are isomorphic. A continuous field of
Hilbert spaces $\rrr{F}_{\Gamma,X}$ is said to be trivial (resp.
locally trivial) if $\bbb{E}(\rrr{F}_{\Gamma,X})$ is trivial (resp.
locally trivial).

\begin{propos}[\cite{fell1} \cite{ dix-dou}]\label{findim}
Let $\rrr{F}_{\Gamma,X}$ be a continuous field of Hilbert spaces
over the compact Hausdorff space $X$ and
$\bbb{E}(\rrr{F}_{\Gamma,X})$ the associated Hilbert bundle. Then:
\begin{enumerate}
\item[{\upshape (i)}] if  $\rrr{F}_{\Gamma,X}$ is separable and $X$ is second-countable (or equivalently metrizable) then the topology defined on the total space $\bbb{E}(\rrr{F})$ is second-countable;
\item[{\upshape (ii)}] if $\text{{\upshape dim}}\ \sss{H}(x)=\aleph_0$ for all $x\in X$ and if $X$ is a finite dimensional
manifold then the Hilbert bundle $\bbb{E}(\rrr{F}_{\Gamma,X})$ is
trivial;
\item[{\upshape (iii)}] if $\text{{\upshape dim}}\ \sss{H}(x)=q<+\infty$ for all $x\in X$ then the Hilbert bundle $\bbb{E}(\rrr{F}_{\Gamma,X})$ is a Hermitian vector bundle with typical fiber $\C^q$.
\end{enumerate}
 \end{propos}

For the proof of (i) one can see \cite{fell1} (Chapter II,
Proposition 13.21). The proof of  (ii) is in \cite{dix-dou} (Theorem
5). As for the proof of (iii), we recall that a Hilbert bundle has
the (stronger) structure of a vector bundle whenever the local
triviality and the continuity of the transition functions (see
\cite{lang2} Chapter III) hold true. However, if the fibers are
finite dimensional then the continuity of the transition functions
follows from the existence of the local trivializations (see
\cite{lang2} Chapter III, Proposition 1), hence one needs only to
prove the local triviality. The latter follows from standard
arguments, as in the final remark of Section 1 of \cite{dix-dou}.

\subsection*{Algebraic viewpoint}
Roughly speaking, a continuous field of Hilbert spaces is an
\virg{analytic object} while a Hilbert bundle is a \virg{geometric
object}. There is also a third point of view which is of algebraic
nature. We introduce an \virg{algebraic object} which encodes all
the relevant properties of the set of continuous vector fields (or
continuous sections).
\begin{defi}[Algebraic viewpoint: Hilbert module]\label{Cmod}
A {\upshape (left) pre-$C^\ast$-module} over a commutative unital
$C^\ast$-algebra $\sss{A}$ is a complex vector space $\Omega_0$ that
is also a (left) $\sss{A}$-module endowed with a  pairing $\{\cdot ;\cdot
\}:\Omega_0 \times \Omega_0 \to\sss{A}$ satisfying, for
$\sigma,\varrho,\varsigma\in\Omega_0$ and for $a\in\sss{A}$ the
following requirements:
\begin{enumerate}
\item[{\upshape a)}] $\{\sigma;\varrho+\varsigma\}=\{\sigma;\varrho\}+\{\sigma;\varsigma\}$;
\item[{\upshape b)}] $\{\sigma;a\varrho\}=a\{\sigma;\varrho\}$;
\item[{\upshape c)}] $\{\sigma;\varrho\}^\ast=\{\varrho;\sigma\}$;
\item[{\upshape d)}]  $\{\sigma;\sigma\}>0$ if $\sigma\neq0$.
\end{enumerate}
The map $|||\cdot |||:\Omega_0 \to[0,+\infty)$ defined  by
$|||\sigma|||:=\sqrt{\|\{\sigma;\sigma\}\|_\sss{A}}$ is a norm in
$\Omega_0$. The completion $\Omega$ of $\Omega_0$ with respect to
the norm $|||\cdot |||$ is called {\upshape (left) $C^\ast$-module}
or {\upshape Hilbert module} over $\sss{A}$.
\end{defi}
\begin{propos}[Equivalence between algebraic and analytic viewpoint \cite{dix-dou}]\label{vecmod}
Let $\rrr{F}_{\Gamma,X}$ be a continuous field of Hilbert  spaces
over the compact Hausdorff space $X$. The set of continuous vector
fields $\Gamma$ has the structure of a Hilbert module over $C(X)$.
Conversely, any Hilbert module over $C(X)$ defines a continuous
field of Hilbert spaces. This correspondence is one-to-one.
\end{propos}

\Proof We shortly sketch the proof, see \cite{dix-dou} (Section 3)
for details. To prove the first part of the statement one observes
that for all pairs of continuous vector fields
$\sigma(\cdot),\varrho(\cdot)\in\Gamma$ the pairing $\{\cdot ;\cdot
\}:\Gamma\times\Gamma\to C(X)$ defined fiberwise by the inner
product, i.e. by posing
$\{\sigma;\varrho\}(x):=(\sigma(x);\varrho(x))_x$, satisfies
Definition \ref{Cmod}. The norm is defined by
$|||\sigma|||:=\sup_{x\in X}\|\sigma(x)\|_x$ and $\Gamma$ is closed
with respect to this norm in view of the property of locally uniform
closure.

Conversely let $\Omega$ be a $C^\ast$-module over $C(X)$. For all
$x\in X$ define a pre-Hilbert structure on $\Omega$ by
$(\sigma;\varrho)_x:=\{\sigma;\varrho\}(x)$. The set
$\sss{I}_x:=\{\sigma\in\Omega\ :\ \{\sigma;\sigma\}(x)=0\}$ is a
linear subspace of $\Omega$. On the quotient space
$\Omega/\sss{I}_x$ the inner product $(\ ;\ )_x$  is a  positive
definite sesqulinear form and we denote by $\sss{H}(x)$ the related
Hilbert space. The collection $\{ \sss{H}(x): x \in X\}$ defines a
field of Hilbert  spaces $\rrr{F}(\Omega)= \prod_{x \in X}
\sss{H}(x)$. For all $\sigma\in\Omega$  the canonical projection
$\Omega\ni
\sigma\stackrel{\jmath_x}{\longmapsto}\sigma(x)\ni\Omega/\sss{I}_x$
defines a vector field $\sigma(\cdot)\in\rrr{F}(\Omega)$. It is easy
to check that the map $\Omega\ni
\sigma\stackrel{\Gamma}{\longmapsto}\sigma(\cdot)\ni\rrr{F}(\Omega)$
is injective. We denote by $\Gamma(\Omega)$ the  image of $\Omega$
in $\rrr{F}(\Omega)$.
 The family  $\Gamma(\Omega)$ defines a continuous structure on $\rrr{F}(\Omega)$. Indeed $\{\sigma(x)\ :\ \sigma(\cdot)\in\Gamma(\Omega)\}=\Omega/\sss{I}_x$ is dense in $\sss{H}(x)$ and $\|\sigma(x)\|^2_x=\{\sigma;\sigma\}(x)$ is continuous. Finally locally uniform closure of $\Gamma(\Omega)$ follows from the closure of $\Omega$ with respect to the norm
$|||\sigma|||:=\sup_{x\in X}\sqrt{\{\sigma;\sigma\}(x)}$ and the
existence of  a partition of the unit subordinate to a   finite
cover of $X$ (since $X$ is compact). \CVD

\subsection*{The Hilbert bundle emerging from the Bloch-Floquet decomposition}

Before proceeding with our analysis, it is useful to summarize in the following diagram the relations between the
algebraic, the analytic and the geometric descriptions.
\begin{equation}\label{dia}
\xymatrix{
&\boxed{\txt{Continuous field\\ of\\ Hilbert spaces\\ $\rrr{F}_{\Gamma,X}$}}\ar@{<->}[dl]_{\mathsf{A}}\ar@{<->}[dr]^{\mathsf{B}}&\\
\ar@{}[drr] |{\txt{fibers of finite dimension $q$}}
 \boxed{\txt{Hilbert bundle\\ $\bbb{E}_{\pi,X}$}}\ar@{->}[d]_{\mathsf{D}}\ar@{<->}[rr]^{\mathsf{C}}& & \boxed{\txt{$C(X)$-module\\ $\Omega$  }}\ar@{->}[d]^{\mathsf{E}}\\
 \boxed{\txt{rank-$q$\\ Hermitian\\ vector bundle  }}
\ar@{<->}[rr]^{\mathsf{F}}& & \boxed{\txt{projective\\
finitely generated\\
$C(X)$-module}}
}
\end{equation}
Arrows $\mathsf{A}$ and $\mathsf{B}$ summarize the content of Propositions
\ref{vecbund} and \ref{vecmod} respectively, arrow $\mathsf{D}$ corresponds
to point (iii) of Proposition \ref{findim}, and  arrow $\mathsf{E}$ follows by
Proposition 53 in \cite{landi}. Arrow $\mathsf{F}$ corresponds to the
remarkable Serre-Swan Theorem (see \cite{landi} Proposition 21), so
arrow $\mathsf{C}$ can be interpreted as a generalization of the Serre-Swan
Theorem.

\bigskip

Coming back to our original problem, let
$\{\sss{H},\rrr{A},\rrr{S}\}$ be a \triple\ with $\sss{H}$ a
separable Hilbert space and $\rrr{S}$ a $\Z^N$-algebra with
generators $\{U_1,\ldots,U_N\}$ and wandering system $\{\psi_k\}_{k
\in \I}$. The Bloch-Floquet decomposition (Theorem \ref{BFspectral})
ensures the existence of a unitary map $\sss{U}_\rrr{S}$, which maps
$\sss{H}$ into the direct integral
$\rrr{K}:=\int^\oplus_{\T^N}\sss{K}(t)\ dz(t)$. Let
$\rrr{F}:=\prod_{t\in\T^N}\sss{K}(t)$ be the corresponding field of
Hilbert spaces. The space $\rrr{K}$ is a subset of $\rrr{F}$ which
has the structure of a Hilbert space and whose elements can be seen
as $L^2$-sections of a \virg{pseudo-Hilbert bundle}
$\bbb{E}(\rrr{F}):=\bigsqcup_{t\in \T^N}\sss{K}(t)$. This justifies
the use of the notation $\rrr{K}=\Gamma_{L^2}(\bbb{E})$.

To obtain a topological decomposition,  we need to know \emph{a
priori} how to select  a continuous structure $\Gamma\subset\rrr{K}$
for the field of Hilbert spaces $\rrr{F}$. In view of Proposition
\ref{vecbund}, this procedure is equivalent to selecting \emph{a
priori} the family of the continuous section $\Gamma(\bbb{E})$ of
the Hilbert bundle $\bbb{E}$ inside the Hilbert space of the
$L^2$-sections $\Gamma_{L^2}(\bbb{E})$. We can use the generalized
Bloch-Floquet transform to push back this problem at the level of
the original Hilbert space $\sss{H}$ and to adopt the algebraic
viewpoint. With this change of perspective the new, but equivalent,
question which we need to answer is: does the \triple\
$\{\sss{H},\rrr{A},\rrr{S}\}$ select a Hilbert module over $C(\T^N)$
inside the Hilbert space $\sss{H}$?

Generalizing an idea of \cite{gruber}, we can use the  transform $\sss{U}_\rrr{S}$ and the notion of wandering nuclear space $\Phi$ to provide a positive answer. The core of our analysis is the following result.
\begin{propos}\label{propHilmod}
 Let $\rrr{S}$ be a $\Z^N$-algebra in the separable Hilbert space
$\sss{H}$ with generators $\{U_1,\ldots,U_N\}$, wandering system
$\{\psi_k\}_{k \in \I}$ and wandering nuclear space $\Phi$. Let
$\rrr{K}$ be the direct integral defined by the Bloch-Floquet
transform $\sss{U}_\rrr{S}:\sss{H}\to\rrr{K}$.  Then the
Bloch-Floquet transform endowes $\Phi$ with the structure of a
(left) pre-$C^\ast$-module over $C(\T^N)$. Let $\Omega_\rrr{S}$ be
the completion of $\Phi$ with respect to the $C^\ast$-module norm.
Then $\Omega_\rrr{S}$ is a Hilbert module over $C(\T^N)$ such that
$\Omega_\rrr{S}\subset\sss{H}$.
\end{propos}

\Proof The set $\Phi$ is a complex vector space which can be endowed
with the structure of a $C(\T^N)$-module by means of the Gel'fand
isomorphism. For any $f\in C(\T^N)$ and $\varphi\in\Phi$ we define
the \emph{(left) module product} $\star$ by
\begin{equation}\label{eq_star_prod}
C(\T^N)\times\Phi\ni(f,\varphi)\longmapsto
f\star\varphi:=A_f\varphi\in\Phi
\end{equation}
where $A_f\in\rrr{S}$ is the operator associated with $f\in
C(\T^N)$. The product is well defined since $\Phi$ is
$\rrr{S}$-invariant by construction. The Bloch-Floquet transform
allows us also to endow $\Phi$ with a pairing
$\{\cdot;\cdot\}:\Phi\times\Phi\to C(\T^N)$. Indeed, for any pair
$\varphi,\phi\in\Phi$ and for all $t\in\T^n$ we define a
sesquilinear form
\begin{equation}\label{eq_pair}
\Phi\times\Phi\ni(\varphi,\phi)\longmapsto\{\varphi,\phi\}(t):=\left((\sss{U}_\rrr{S}\varphi)(t);(\sss{U}_\rrr{S}\phi)(t)\right)_t\in
\C.
\end{equation}
Moreover $\{\varphi,\phi\}(t)$ is
 a continuous function of $t$.
Indeed  $\varphi,\phi\in\Phi$ means that  $\varphi$ and $\phi$ are finite linear combinations of the
vectors $U^a\psi_k$ and from equation \eqref{eq6} and the orthonormality of the fundamental vector fields $\zeta_k(\cdot)$ it follows that
 $\{\varphi,\phi\}(t)$ consists of a finite linear combination of the exponentials $\expo{it_1},\ldots,\expo{it_N}$.

Endowed with the operations \eqref{eq_star_prod} and \eqref{eq_pair}, the space $\Phi$ becomes a (left) pre-$C^\ast$-module
 over $C(\T^N)$.
The Hilbert module $\Omega_\rrr{S}$ is defined to be the completion
of $\Phi$ with respect to the norm
\begin{equation}\label{eqcro1}
|||\varphi|||^2:=\sup_{t\in\T^N}\|(\sss{U}_\rrr{S}\varphi)(t)\|^2_t=\sup_{t\in\T^N}\left(\sum_{k
\in \I}^\text{fin}|f_{\varphi;k}(t)|^2\right)
\end{equation}
 according to the notation in the proof of Theorem \ref{BFspectral}. Let $\{\varphi_n\}_{n\in\N}$ be a sequence
 in $\Phi$ which is Cauchy with respect to the norm $|||\cdot|||$. From \eqref{eqcro1}, the unitarity
 of $\sss{U}_\rrr{S}$ and the normalization of the Haar measure $dz$ on $\T^N$ it follows that
 $\|\varphi_n-\varphi_m\|_\sss{H}\leqslant|||\varphi_n-\varphi_m|||$, hence $\{\varphi_n\}_{n\in\N}$ is also Cauchy
 with respect to the norm $\|\cdot\|_\sss{H}$, so the limit $\varphi_n\to\varphi$ is an element of $\sss{H}$.
\CVD
Once   the Hilbert module $\Omega_\rrr{S}$ is selected, we can use
it to define a continuous field of Hilbert spaces as explained in
Proposition \ref{vecmod}. It is easy to convince oneself that the
abstract construction proposed in Proposition \ref{vecmod} is
concretely implemented by the generalized Bloch-Floquet transform
$\sss{U}_\rrr{S}$. Then the set of vector fields
$\Gamma_\rrr{S}:=\sss{U}_\rrr{S}(\Omega_\rrr{S})$ defines a
continuous structure on the field of Hilbert spaces
$\rrr{F}:=\prod_{t\in\T^N}\sss{K}(t)$ and, in view of Proposition
\ref{vecbund}, a Hilbert bundle over the base manifold $\T^N$. This
Hilbert bundle, which we will denote by $\bbb{E}_\rrr{S}$, is the
set $\bigsqcup_{t\in \T^N}\sss{K}(t)$ equipped by the topology
prescribed by the set of the continuous sections $\Gamma_\rrr{S}$.
The structure of $\bbb{E}_\rrr{S}$ depends only on the equivalence
class of the \triple\ $\{\sss{H},\rrr{A},\rrr{S}\}$ and we will
refer to it as the \emph{Bloch-Floquet Hilbert bundle}.

\begin{teo}[Emerging geometric structure]\label{emerging_geo}
Let $\rrr{S}$ be a $\Z^N$-algebra in the separable Hilbert space
$\sss{H}$ with generators $\{U_1,\ldots,U_N\}$, wandering system
$\{\psi_k\}_{k \in \I}$ and wandering nuclear space $\Phi$. Let
$\rrr{K}$ be the direct integral defined by the Bloch-Floquet
transform $\sss{U}_\rrr{S}:\sss{H}\to\rrr{K}$ and
$\Omega_\rrr{S}\subset\sss{H}$ the Hilbert module over $C(\T^N)$
defined in Proposition \ref{propHilmod}. Then:
\begin{enumerate}
\item[{\upshape (i)}] the  family of vector fields $\sss{U}_\rrr{S}(\Omega_\rrr{S})=:\Gamma_\rrr{S}$ defines a continuous structure on  $\rrr{F} = \Pi_{t\in \T^N} \sss{K}(t)$ which realizes the correspondence stated in Proposition \ref{vecmod};
\item[{\upshape (ii)}] the Bloch-Floquet Hilbert bundle $\bbb{E}_\rrr{S}$, defined  by $\Gamma_\rrr{S}$ according to Proposition \ref{vecbund}, depends only on the equivalence class of the \triple\   $\{\sss{H},\rrr{A},\rrr{S}\}$.
\end{enumerate}
\end{teo}
\Proof To prove (i) let $\sss{I}_t:=\{\varphi\in\Phi\ :\
\left((\sss{U}_\rrr{S}\varphi)(t);(\sss{U}_\rrr{S}\varphi)(t)\right)_t=0\}$.
The space $\Phi/ \sss{I}_t$ is a pre-Hilbert space with respect to
the scalar product induced by $\left.\sss{U}_\rrr{S}\right|_t$. The
map $\left.\sss{U}_\rrr{S}\right|_t:\Phi/ \sss{I}_t\to\sss{K}(t)$ is
obviously isometric and so can be extended to a linear isometry from
the norm-closure of $\Phi/ \sss{I}_t$ into $\sss{K}(t)$. The map
$\left.\sss{U}_\rrr{S}\right|_t$ is also surjective, indeed
$\sss{K}(t)$ is generated by the orthonormal basis
$\{\zeta_k(t)\}_{k \in \I}$ and
${\left.\sss{U}_\rrr{S}\right|_t}^{-1}\zeta_k(t)=\psi_k\in\Phi/
\sss{I}_t$. Then the fiber Hilbert spaces appearing in the proof of
Proposition \ref{vecmod} coincide, up to a unitary equivalence, with
the fiber Hilbert spaces $\sss{K}(t)$ obtained through the
Bloch-Floquet decomposition. Moreover the Bloch-Floquet transform
acts as the map defined in the proof of Proposition \ref{vecmod},
which sends any element of the Hilbert module $\Phi$ to a continuous
section of $\rrr{F}$.

To prove (ii) let $\{\sss{H}_1,\rrr{A}_1,\rrr{S}_1\}$ and
$\{\sss{H}_2,\rrr{A}_2,\rrr{S}_2\}$ be two \triple s related by a
unitary map $U:\sss{H}_1\to\sss{H}_2$. If $\rrr{S}_1$ is a
$\Z^N$-algebra in $\sss{H}_1$ then also $\rrr{S}_2=U\rrr{S}_1U^{-1}$
is a $\Z^N$-algebra in $\sss{H}_2$ and if
$\{\psi_k\}_{k \in \I}\subset\sss{H}_1$ is a wandering system for
$\rrr{S}_1$ then
$\{\widetilde{\psi}_k:=U\psi_k\}_{k \in \I}\subset\sss{H}_2$ is a
wandering system for $\rrr{S}_2$ (with the same cardinality). The
two wandering nuclear spaces $\Phi_1\subset\sss{H}_1$ and
$\Phi_2\subset\sss{H}_2$ are related by $\Phi_2=U\Phi_1$. Let
$\sss{U}_{\rrr{S}_1}:\sss{H}_1\to\rrr{H}_1$ and
$\sss{U}_{\rrr{S}_2}:\sss{H}_2\to\rrr{H}_2$ be the two generalized
Bloch-Floquet transforms defined by the two equivalent \triple s.
From the explicit expression of $\sss{U}_{\rrr{S}_2}$ and
${\sss{U}_{\rrr{S}_1}}^{-1}$, and in accordance with Corollary
\ref{appcor1}, one argues that $\sss{U}_{\rrr{S}_2}\circ
U\circ{\sss{U}_{\rrr{S}_1}}^{-1}=:W(\cdot)$ is a decomposable
unitary which is well defined for all $t$.
 Let $\varphi,\phi\in\Phi_1$ then
\begin{align*}
\{\varphi;\phi\}_1(t):&=\left((\sss{U}_{\rrr{S}_1}\varphi)(t);(\sss{U}_{\rrr{S}_1}\phi)(t)\right)_t=\left(W(t)(\sss{U}_{\rrr{S}_1}\varphi)(t);W(t)(\sss{U}_{\rrr{S}_1}\phi)(t)\right)_t\\
&=\left((\sss{U}_{\rrr{S}_2}U\varphi)(t);(\sss{U}_{\rrr{S}_2}U\phi)(t)\right)_t=\left((\sss{U}_{\rrr{S}_2}\widetilde{\varphi})(t);(\sss{U}_{\rrr{S}_2}\widetilde{\phi})(t)\right)_t=:\{\widetilde{\varphi};\widetilde{\phi}\}_2(t)
\end{align*}
where $\widetilde{\varphi}:=U\varphi$ and $\widetilde{\phi}:=U\phi$
are elements of $\Phi_2$. This equation shows that $\Phi_1$ and
$\Phi_2$ have the same $C(\T^N)$-module structure and so define the
same abstract Hilbert module over $C(\T^N)$. The claim follows from
the generalization of the Serre-Swan Theorem summarized by arrow
$\mathsf{C}$ in \eqref{dia}. \CVD
\begin{rk}
With a proof similar to that of point (ii) of Theorem
\ref{emerging_geo}, one deduces also that the Bloch-Floquet-Hilbert
bundle $\bbb{E}_\rrr{S}$ does not depend on the choice of two
unitarily (or antiunitarily) equivalent commutative
$C^\ast$-algebras $\rrr{S}_1$ and $\rrr{S}_2$ inside $\rrr{A}'$.
Indeed, also in this case the  abstract Hilbert module structure
induced by the two Bloch-Floquet transforms $\sss{U}_{\rrr{S}_1}$
and $\sss{U}_{\rrr{S}_2}$ is the same. \hfill
$\blacklozenge\lozenge$
\end{rk}

After defining the topology of the Bloch-Floquet Hilbert bundle, it is natural
to deduce  more information about its structure. An interesting property arises from the cardinality of the wandering
system, which depends only on  the \triple\ (see Corollary
\ref{corocard}).
\begin{corol}\label{corhilvec}
  The Hilbert bundle $\bbb{E}_\rrr{S}$ over the torus $\T^N$ defined by the continuous structure
$\Gamma_\rrr{S}$  is trivial if the cardinality of the wandering
system is $\aleph_0$, and is a rank-$q$ Hermitian vector bundle if
the cardinality of the wandering system is $q$. In the latter case
the transition functions of the vector bundle can be expressed in
terms of the fundamental orthonormal frame
$\{\zeta_k(\cdot):=(\sss{U}_\rrr{S}\psi_k)\}_{k=1,\ldots,q}$.
\end{corol}

{\subsection*{Decomposition of the observables and endomorphism sections}}

According to Theorem \ref{BFspectral}, the Bloch-Floquet transform
\eqref{app13} provides a concrete realization for the unitary map
($\rrr{S}$-Fourier transform) whose existence is claimed by  von
Neumann's complete spectral theorem. Point (ii) of Theorem
\ref{compspec} implies that under the Bloch-Floquet transform  any
$O\in\rrr{A}$ is mapped into a decomposable operator on the direct
integral $\int_{\T^N}^\oplus\sss{K}(t)\ dz(t)$ , i.e.
$\sss{U}_\rrr{S} \, O \, {\sss{U}_\rrr{S}}^{-1}=:O(\cdot):t\mapsto
O(t)\in\bbb{B}(\sss{K}(t))$ with $O(\cdot)$ weakly mesurable.

The natural question which arises is the following: there exists any
topological structure in the $C^\ast$-algebra $\rrr{A}$ compatible
with the Bloch-Floquet Hilbert bundle which emerges from the
Bloch-Floquet transform?  In order to answer  this question, we first
analyze the nature of the linear maps which preserve the (Hilbert
module) structure of the set of the continuous sections.

\begin{defi}[Hilbert module endomorphism]\label{Def adjointable}
Let $\Omega$ be a (left) Hilbert module over the commutative unital
$C^\ast$-algebra $\sss{A}$. An {\upshape endomorphism} of \,
$\Omega$ is an $\sss{A}$-linear map $O:\Omega\to\Omega$ which is
\emph{adjointable}, i.e. there exists a map $O^\dag:\Omega\to\Omega$
such that $\{\sigma;O\rho\}=\{O^\dag\sigma;\rho\}$ for all
$\sigma,\rho\in\Omega$.
 We denote by $\text{{\upshape End}}_{\sss{A}}(\Omega)$ the set of all  endomorphisms of $\Omega$.
\end{defi}
As proven  in \cite{bon} (Section 2.5) or \cite{landi} (Appendix A),
if $O\in\text{{\upshape End}}_{\sss{A}}(\Omega)$, then  $O^\dag\in\text{{\upshape End}}_{\sss{A}}(\Omega)$ and
$\dag$ is an involution over $\text{{\upshape
End}}_{\sss{A}}(\Omega)$. Moreover, $\text{{\upshape
End}}_{\sss{A}}(\Omega)$ endowed with the \emph{endomorphism norm}
\begin{equation}\label{eqnew2}
 \|O\|_{\text{End}(\Omega)}:=\text{sup}\{|||O(\sigma)|||\ :\ |||\sigma|||\leqslant1\}
\end{equation}
 becomes a $C^\ast$-algebra (of bounded operators). For any $\sigma,\rho\in\Omega$ one defines the \emph{rank-1 endomorphism} $|\sigma\}\{\rho|\in\text{{\upshape End}}_{\sss{A}}(\Omega)$ by
$|\sigma\}\{\rho|(\varsigma):=\{\rho;\varsigma\}\ \sigma$ for all
$\varsigma\in\Omega$. The adjoint of $|\sigma\}\{\rho|$ is given by
$|\rho\}\{\sigma|$. The linear span of the rank-1 endomorphisms is a
selfadjoint two-sided ideal of $\text{{\upshape
End}}_{\sss{A}}(\Omega)$ (\emph{finite rank endomorphisms}) and its
(operator) norm closure is denoted by $\text{{\upshape
End}}^0_{\sss{A}}(\Omega)$. The elements of the latter are called
\emph{compact endomorphisms} of $\Omega$. Since $\text{{\upshape
End}}^0_{\sss{A}}(\Omega)$ is an \emph{essential ideal} of
$\text{{\upshape End}}_{\sss{A}}(\Omega)$, it follows that
$\text{{\upshape End}}^0_{\sss{A}}(\Omega)=\text{{\upshape
End}}_{\sss{A}}(\Omega)$ if and only if
$\num{1}_{\Omega}\in\text{{\upshape End}}^0_{\sss{A}}(\Omega)$.

A remarkable result which emerges from  the above theory is the
characterization of the compact endomorphisms of the $C(X)$ Hilbert
module $\Gamma(\sss{E})$ of the continuous sections of a rank-$q$
Hermitian vector bundle.

\begin{propos}\label{propnew3}
Let $\bbb{E}\to X$ be a rank-$q$ Hermitian vector bundle over the compact Hausdorff space $X$ and let $\Gamma(\bbb{E})$ be the  Hilbert module over $C(X)$ of its continuous sections. Then
\begin{equation}\label{eq_loc_iso}
\text{{\upshape End}}^0_{C(X)}(\Gamma(\bbb{E}))=\text{{\upshape End}}_{C(X)}(\Gamma(\bbb{E}))\simeq\Gamma(\text{{\upshape End}}(\bbb{E}))
\end{equation}
where $\Gamma(\text{{\upshape End}}(\bbb{E}))$ denotes the continuous sections of the  vector bundle $\text{{\upshape End}}(\bbb{E})\to X$. The {\upshape localization isomorphism} appearing in right-hand side of \eqref{eq_loc_iso} preserves the composition and the structure of $C(X)$-module.
\end{propos}

The proof is a consequence of the  Serre-Swan Theorem (see
\cite{bon}, Theorems 2.10 and 3.8) and of Proposition 3.2 in
\cite{bon}.

In Proposition \ref{propHilmod} we proved that the Gel'fand
isomorphism and the  Bloch-Floquet transform
 equip the wandering nuclear space $\Phi$ with the structure of a (left) pre-$C^\ast$-module over $C(\T^N)$
by means of the (left) product $\star$ defined by
\eqref{eq_star_prod} and  the pairing $\{\ ;\ \}$ defined by
\eqref{eq_pair}. The closure of $\Phi$ with respect to the module
norm defines a Hilbert module over $C(\T^N)$ denoted by
$\Omega_\rrr{S}\subset\sss{H}$. In this description, what is the
role played by $\rrr{A}$? Is it possible, at least under some
condition, to interpret the elements of $\rrr{A}$ as endomorphism of
the Hilbert module $\Omega_\rrr{S}$? One could try to answer  these
questions by observing that  for any $O\in\rrr{A}$, any
$A_f\in\rrr{S}$ and any $\varphi\in\Omega_\rrr{S}$ one has
  that $O(f\star\varphi):=OA_f\varphi=A_fO\varphi$. The latter might be interpreted as $f\star O(\varphi)$, implying the $C(\T^N)$-linearity of  $O\in\rrr{A}$ as operator on $\Omega_\rrr{S}$. However it may happen that $O\varphi\notin \Omega_\rrr{S}$ which implies that $O$ can not define an endomorphism of $\Omega_\rrr{S}$. Everything works properly if one considers only elements in the subalgebra $\rrr{A}^0\subset\rrr{A}$ defined by
\begin{equation}
\rrr{A}^0:=\{O\in\rrr{A}\ :\ O:\Omega_\rrr{S}\to\Omega_\rrr{S}\}.
\end{equation}
\begin{propos}\label{propnew1}
 Let $\Omega_\rrr{S}$ be the Hilbert module over $C(\T^N)$
defined by means of the Bloch-Floquet transform according to
Proposition \ref{propHilmod}. Let $\rrr{A}^0_\text{{\upshape s.a.}}$
be  the $C^\ast$-subalgebra of $\rrr{A}$ defined by
$\rrr{A}^0_\text{{\upshape s.a.}}:=\{O\in\rrr{A}\,:\,
O,O^\dag\in\rrr{A}^0\}$ (self-adjoint part of $\rrr{A}^0$). Then
$\rrr{A}^0_\text{{\upshape s.a.}}\subset\text{\upshape
End}_{C(\T^N)}(\Omega_\rrr{S})$.
\end{propos}

\Proof Let $O\in\rrr{A}^0_\text{{\upshape s.a.}}$. By definition $O$
is a linear map from $\Omega_\rrr{S}$ to itself;  it is also
$C(\T^N)$-linear since $O(f\star\varphi)=O A_f\varphi=A_fO\varphi$
as mentioned. We need to prove that $O$ is bounded with respect to
the endomorphism norm \eqref{eqnew2}. From the definition
\eqref{eqcro1} of the module norm $|||\cdot|||$   it follows that
$$
|||O
\varphi|||=\text{sup}_{t\in\T^N}\|(\sss{U}_\rrr{S}O\varphi)(t)\|_t=\text{sup}_{t\in\T^N}\|\pi_t(O)(\sss{U}_\rrr{S}\varphi)(t)\|_t\leqslant
\|O\|_{\bbb{B}(\sss{H})}\ |||\varphi|||
$$
where $\pi_t(O):=\left.\sss{U}_\rrr{S}\right|_t\ O\
{\left.\sss{U}_\rrr{S}\right|_t}^{-1}$ defines a representation of
the $C^\ast$-algebra $\rrr{A}$ on the fiber Hilbert space
$\sss{K}(t)$ and
$\|\pi_t(O)\|_{\bbb{B}(\sss{K}(t))}\leqslant\|O\|_{\bbb{B}(\sss{H})}$
since any $C^\ast$ representation decreases the norm. Thus
$\|O\|_{\text{End}(\Omega_\rrr{S})}\leqslant\|O\|_{\bbb{B}(\sss{H})}$,
therefore $O$ defines a continuous $C(\T^N)$-linear map from
$\Omega_\rrr{S}$ to itself. To prove that
$O\in\text{End}_{C(\T^N)}(\Omega_\rrr{S})$ we must show that $O$ is
adjointable, which follows from the definition of
$\rrr{A}^0_\text{{\upshape s.a.}}$. \CVD

It is of particular interest to specialize the previous result to
the case of a finite wandering system.
\begin{teo}[Bloch-Floquet endomorphism bundle]\label{teonewnew}
Let $\{\sss{H},\rrr{A},\rrr{S}\}$ be a \triple\ where $\rrr{S}$ is a $\Z^N$-algebra with generators $\{U_1,\ldots,U_N\}$ and  wandering system $\{\psi_1,\ldots,\psi_q\}$ of finite cardinality. Then:
\begin{enumerate}
\item[{\upshape (i)}] $\rrr{A}^0_\text{{\upshape s.a.}}=\rrr{A}^0$;
\item[{\upshape (ii)}] $\sss{U}_\rrr{S}\ \rrr{A}^0\ {\sss{U}_\rrr{S}}^{-1}\subseteq\Gamma(\text{{\upshape End}}(\bbb{E}_\rrr{S}))$ where $\bbb{E}_\rrr{S}\to\T^N$ is the rank $q$
Bloch-Floquet vector bundle defined in Corollary \ref{corhilvec}.
\end{enumerate}
\end{teo}
\Proof
To prove (i) let $O\in\rrr{A}^0$ and observe that if
 $O\psi_k=\sum_{h=1}^q\sum_{b\in\Z^N}\alpha_{h,b}^{(k)}\ U^b\psi_h$ then $O^\dag\psi_k=\sum_{h=1}^q\sum_{b\in\Z^N}\overline{\alpha}_{k,b}^{(h)}\ U^{-b}\psi_h$.
Since $O\psi_k\in\Omega_\rrr{S}$, then
$f^{(k)}_h(t):=\sum_{b\in\Z^N}\alpha_{h,b}^{(k)}\ z^b(t)$ is a
continuous function on $\T^N$ and
$$
|||O^\dag\psi_k|||^2
=\text{sup}_{t\in\T^N}\left(\sum_{h=1}^q|f^{(h)}_k(t)|^2\right)<+\infty.
$$
Then $O^\dag\psi_k\in\Omega_\rrr{S}$ for all $k=1,\ldots,q$. Since
$O^\dag (U^b\psi_k)=U^b(O^\dag\psi_k)\in\Omega_\rrr{S}$ for all
$b\in\Z^N$ it follows that also $O^\dag\in\rrr{A}^0$. Point (ii) is
an immediate consequence of Proposition \ref{propnew1}, Corollary
\ref{corhilvec} and Proposition \ref{propnew3}. \CVD

\begin{ex}[\emph{Mathieu-like Hamiltonians, part four}]
It is immediate to check that both $\rrr{u}$ and $\rrr{v}$ preserve
the wandering nuclear space $\Phi$, so that the full
$C^\ast$-algebra $\rrr{A}_\text{M}^{\nicefrac{p}{q}}$ consists of
endomorphisms for the Hilbert module realized by means of the
Bloch-Floquet transform $\sss{U}_{\rrr{S}^q_\text{M}}$. Theorem
\ref{teonewnew} claims that $\sss{U}_{\rrr{S}^q_\text{M}}$ maps
$\rrr{A}_\text{M}^{\nicefrac{p}{q}}$ into a subalgebra of the
endomorphisms  of the trivial bundle $\T\times\C^q\to\T$. The
matrices $\rrr{u}(t)$ and $\rrr{v}(t)$ in Example \ref{exnew1}
define the representation of the generators as elements of
$\Gamma(\text{End}(\T\times\C^q))\simeq
C(\T)\otimes\text{Mat}_q(\C)$. \hfill
$\blacktriangleleft\vartriangleright$
\end{ex}


\appendix
\section{Gel'fand theory, joint spectrum and basic measures}\label{secgelf}
Let $\rrr{A}$ be a unital (not necessarily commutative)
$C^\ast$-algebra and $\rrr{A}^\times$ the \emph{group of the
invertible elements} of $\rrr{A}$. The \emph{algebraic spectrum} of
$A\in\rrr{A}$ is defined to be  $\sigma_\rrr{A}(A):=\{\lambda\in\C\
:\ (A-\lambda\num{1})\notin\rrr{A}^\times\}$. If $\rrr{A}_0$ is a
non unital $C^\ast$-algebra and
$\imath:\rrr{A}_0\hookrightarrow\rrr{A}$ is the canonical embedding
of $\rrr{A}_0$ in the unital $C^\ast$-algebra $\rrr{A}$ (see
\cite{bra-rob1} Proposition 2.1.5) then one defines
$\sigma_{\rrr{A}_0}(A):=\sigma_{\rrr{A}}(\imath(A))$ for all
$A\in\rrr{A}_0$. If $\rrr{A}$ is unital and
$C^\ast(A)\subset\rrr{A}$ is the unital $C^\ast$-subalgebra
generated algebraically by $A$, its adjoint $A^\dag$ and  $\num{1}$
($=:A^0$ for definition) then
$\sigma_{\rrr{A}}(A)=\sigma_{C^\ast(A)}(A)$ (see \cite{bra-rob1}
Proposition 2.2.7). As a consequence we have that if
$\rrr{A}\subset\bbb{B}(\sss{H})$ is a concrete $C^\ast$-algebra of
operators on the Hilbert space $\sss{H}$ and $A\in\rrr{A}$ then the
algebraic spectrum $\sigma_\rrr{A}(A)$ agrees with the \emph{
Hilbert space spectrum} $\sigma(A):=\{\lambda\in\C\ :\
(A-\lambda\num{1})\notin\text{GL}(\sss{H})\}$ where
$\text{GL}(\sss{H}):=\bbb{B}(\sss{H})^\times$ is the group of the
invertible bounded linear operators on the Hilbert space $\sss{H}$.

Let us denote by $\rrr{S}$ a commutative $C^\ast$-algebra. A \emph{
character} of $\rrr{S}$ is a nonzero homomorphism $x:\rrr{S}\to\C$
(also called pure state). The \emph{Gel'fand spectrum} of $\rrr{S}$,
denoted by $X(\rrr{S})$ or simply by $X$, is the set of all
characters of $\rrr{S}$. The space $X$, endowed with the weak-$\ast$
topology (topology of the pointwise convergence on $\rrr{S}$)
becomes a topological Hausdorff space, which is compact if $\rrr{S}$
is unital and only locally compact otherwise (see \cite{bra-rob1}
Theorem 2.1.11A). If $\rrr{S}$ is \emph{separable} (namely it is
generated algebraically by a countable family of commuting elements)
then the weak-$\ast$ topology in $X$ is metrizable (see
\cite{brezis} Theorem III.25) and if, in addition, $\rrr{S}$ is also
unital then $X$ is compact and metrizable which implies (see
\cite{choq} Proposition 18.3 and Theorem 20.9) that $X$ is
second-countable (has a countable basis of open sets),  separable
(has a countable everywhere dense subset) and complete. Summarizing,
the Gel'fand spectrum of a commutative separable unital
$C^\ast$-algebra has the structure of a \emph{Polish space}
(separable complete metric space).

The \emph{Gel'fand-Na\v{\i}mark Theorem} (see \cite{bra-rob1} Section
2.3.5 or \cite{bon} Section 1.2 or \cite{landi} Section 2.2) states
that there is a canonical isomorphism between any commutative
unital $C^\ast$-algebra $\rrr{S}$ and the commutative
$C^\ast$-algebra $C(X)$ of the continuous complex valued functions
on its spectrum endowed with the norm of the uniform convergence.
The \emph{Gel'fand isomorphism} $C(X)\ni f\stackrel{\bbb{G}}{\mapsto}
A_f\in\rrr{S}$ maps  any continuous $f$ into the unique element  $A_f$
which satisfies the relation $f(x)=x(A_f)$ for all $x\in X$. Then we
can use the continuous functions on $X$ to \virg{label} the elements of
$\rrr{S}$. If $\rrr{S}_0$ is a non-unital commutative
$C^\ast$-algebra then the Gel'fand-Na\v{\i}mark Theorem proves the
isomorphism between $\rrr{S}_0$ and the commutative $C^\ast$-algebra
$C_0(X_0)$ of the continuous complex valued functions vanishing at
infinity on the locally compact space $X_0$ which is the spectrum of
$\rrr{S}_0$.
If $\rrr{S}_0\subset\bbb{B}(\sss{H})$ we define the
\emph{multiplier algebra} (or \emph{idealizer}) of $\rrr{S}_0$ to be
$\rrr{S}:=\{B\in\bbb{B}(\sss{H})\ :\ BA,AB\in \rrr{S}_0\ \ \ \forall
A\in\rrr{S}_0\}$ (see \cite{bon} Definition 1.8 and Lemma 1.9).
Obviously $\rrr{S}$ is a unital $C^\ast$-algebra and the commutativity of $\rrr{S}_0$ implies  the commutativity of $\rrr{S}$.
Moreover $\rrr{S}$ contains $\rrr{S}_0$
as an \emph{essential ideal}. The Gel'fand spectrum $X$ of $\rrr{S}$
corresponds to the \emph{Stone-$\check{\text{C}}$ech
compactification} of the spectrum $X_0$. Since
$C(X)\simeq C_\text{b}(X_0)$, the Gel'fand isomorphism asserts that the
multiplier algebra $\rrr{S}$ can be described as the unital
commutative $C^\ast$-algebra of the bounded continuous functions on
the locally compact space $X_0$  (see \cite{bon} Proposition 1.10).
For every $A_f\in\rrr{S}$ one has that $\sigma_\rrr{S}(A_f)=\{f(x)\
:\  x\in X\}$ (see \cite{horm} Theorem 3.1.6) then $A_f$ is
invertible if and only if $0 < |f(x)|\leqslant\|A_f\|_\rrr{S}$ for
all $x\in X$.

We often consider the relevant case when the unital commutative
$C^\ast$-algebra is \emph{finitely generated}, i.e. when $\rrr{S}$
is generated by a finite family $\{A_1,\ldots,A_N\}$ of commuting
normal elements and the identity $\num{1}$ ($=:A^0_j$ by
definition). Let $f_1,\ldots,f_N$ be the continuous functions which
label the elements of the generating system. The map $X\ni
x\stackrel{\varpi}{\mapsto} (f_1(x),\ldots,f_N(x))\in \C^N$ is a
homeomorphism from the Gel'fand spectrum $X$ to a compact subset of
$\C^N$ called the \emph{joint spectrum} of the generating system
$\{A_1,\ldots,A_N\}$ (see \cite{horm} Theorem 3.1.15). Then, when
$\rrr{S}$ is finitely generated, we can identify the Gel'fand
spectrum $X$ with its homeomorphic image $\varpi(X)$ (the joint
spectrum) which is a compact, generally proper, subset of
$\sigma_\rrr{S}(A_1)\times\ldots\times \sigma_\rrr{S}(A_N)$. When
$\{A_1,\ldots,A_N\}\subset\bbb{B}(\sss{H})$ a necessary and
sufficient condition for $\lambda:=(\lambda_1,\ldots,\lambda_N)$ to
be in $\varpi(X)$ is that there exists a sequence of normalized
vectors $\{\psi_n\}_{n\in\N}$ such that
$\|(A_j-\lambda_j)\psi_n\|\to0$ if $n\to\infty$ for all
$j=1,\ldots,N$ (see \cite{samoil} Proposition 2).

\begin{rk}[\emph{Dual group}]\label{rkdualgp}
The Gel'fand theory has an interesting application to abelian
locally compact groups $\num{G}$. Usually the  \emph{dual group} (or
character group) $\widehat{\num{G}}$ is defined to be the set of all
continuous characters of $\num{G}$, namely the set of all
continuous homomorphisms of $\num{G}$ into the group
$\num{S}^1:=\{z\in\C\ :\ |z|=1\}$. However, to endow
$\widehat{\num{G}}$ with a natural topology it is useful to give an
equivalent definition of dual group. Since $\num{G}$ is locally
compact and abelian there exists a
 unique (up to a multiplicative constant) invariant  \emph{Haar measure}
 on $\num{G}$ denoted by $dg$. The space $L^1(\num{G})$ becomes a commutative Banach $\ast$-algebra, if multiplication is defined  by convolution; it is called the \emph{group algebra} of $\num{G}$.
 If $\num{G}$ is discrete then $L^1(\num{G})$ is unital otherwise $L^1(\num{G})$ has always
 an \emph{approximate unit} (see \cite{rudi} Theorems 1.1.7 and 1.1.8).
 Every $\chi\in\widehat{\num{G}}$
defines a linear multiplicative functional $\widehat{\chi}$ on
$L^1(\num{G})$ by $\widehat{\chi}(f):=\int_\num{G}f(g)\chi(-g)\
d\mu(g)$ for all $f\in L^1(\num{G}) $ (the Fourier transform). This
map defines a one to one correspondence between $\widehat{\num{G}}$
and the Gel'fand spectrum of the algebra $L^1(\num{G})$ (see
\cite{rudi} Theorem 1.2.2). This enables us to consider
$\widehat{\num{G}}$ as the Gel'fand spectrum of $L^1(\num{G})$. When
$\widehat{\num{G}}$ is endowed with the weak-$\ast$ topology with
respect to $L^1(\num{G})$ then it becomes a Hausdorff locally
compact space.
 Moreover $\widehat{\num{G}}$ is compact if $\num{G}$ is discrete
 and it is discrete when $\num{G}$ is compact (see \cite{rudi} Theorem 1.2.5).\hfill $\blacklozenge\lozenge$
\end{rk}

Let $X$ be a compact Polish space and $\ssss{B}(X)$  the  Borel
$\sigma$-algebra generated by the topology of $X$. The pair
$\{X,\ssss{B}(X)\}$ is called \emph{standard Borel space}. A mapping
$\mu:\ssss{B}(X)\to[0,+\infty]$ such that $\mu(\emptyset)=0$ and
$\mu(X)< \infty$, which is additive with respect to the union of
countable families of pairwise disjoint subsets of $X$, is called a
\emph{finite Borel measure}. If $\mu(X)=1$ then we will said that
$\mu$ is a \emph{probability} Borel measure.
 Any Borel measure on a standard Borel space $\{X,\ssss{B}(X)\}$ is \emph{regular},
 i.e. for all $Y\in\ssss{B}(X)$
 one has that $\mu(Y)=\sup\{\mu(K)\, :\, K\subset Y,\ K\ \text{compact}\} =
 \inf\{\mu(O)\,:\, Y\subset O,\ O\ \text{open}\}$.

Let $N$ be the union of all the open sets $O_\alpha\subset X$ such
that $\mu(O_{\alpha})=0$. The closed set $X\setminus N$ is called
the \emph{ support} of  $\mu$. If $\mu$ is a regular Borel measure
then $\mu(N)=0$ and $\mu$ is concentrated on its support.

Let $\rrr{S}$ be a unital commutative $C^\ast$-algebra acting on the
separable Hilbert space $\sss{H}$ with Gel'fand spectrum $X$. For
all pairs $\psi,\varphi\in\sss{H}$ the mapping $C(X)\ni
f\mapsto(\psi;A_f\varphi)_\sss{H}\in\C$ is a continuous linear
functional on $C(X)$; hence the \emph{Riesz-Markov Theorem} (see
\cite{rudi2} Theorem 2.14) implies the existence of a unique regular
(complex) Borel measure $\mu_{\psi,\varphi}$, with finite total
variation, such that $ (\psi;A_f\varphi)_\sss{H}=\int_Xf(x)\
d\mu_{\psi,\varphi}(x) $ for all $f\in C(X)$. We will refer to
$\mu_{\psi,\varphi}$ as a \emph{ spectral measure}. The union of the
supports of the (positive) spectral measures $\mu_{\psi,\psi}$ is
dense, namely for every non-void open set $O\subset X$ there exists
a $\psi\in\sss{H}$ such that $\mu_{\psi,\psi}(O)>0$. A positive
measure $\mu$ on $X$ is said to be \emph{basic} for the unital
$C^\ast$-algebra $\rrr{S}$ if: for every $Y \subset X$, $\mu(Y)= 0$
if and only if  $\mu_{\psi,\psi}(Y)=0$ for every $\psi\in\sss{H}$.
From the definition it follows that: (i) if there exists a basic
measure $\mu$ on $X$, then every other basic measure is
\emph{equivalent} (has the same null sets) to $\mu$; (ii) for all
$\psi,\varphi\in\sss{H}$ the spectral measure $\mu_{\psi,\varphi}$
is \emph{absolutely continuous} with respect to $\mu$, and there
exists a unique element $h_{\psi,\varphi}\in L^1(X)$ (the \emph{
Radon-Nikodym derivative}) such that
$\mu_{\psi,\varphi}=h_{\psi,\varphi}\mu$; (iii) since the union of
the supports of the measures $\mu_{\psi,\psi}$ is dense in $X$, then
the support of a basic measure $\mu$ is all of $X$  (see \cite{dix2}
Part I, Chapter 7). The existence of a basic measure for a
commutative $C^\ast$-algebra $\rrr{S}\subset\bbb{B}(\sss{H})$
follows from general arguments. Indeed the existence of a basic
measure is equivalent to the existence of a cyclic vector $\phi$ for
the commutant $\rrr{S}'$ and the basic measure can be chosen to be
the spectral measure $\mu_{\phi,\phi}$ (see \cite{dix2} Part I,
Chapter 7, Proposition 3). Since a vector $\phi$ is cyclic for
$\rrr{S}'$ if and only if it is separating for the commutative von
Neumann algebra $\rrr{S}''\supset\rrr{S}$, and since any commutative
von Neumann algebra of operators on a separable Hilbert space has a
separating vector, it follows that any commutative unital
$C^\ast$-algebra $\rrr{S}$ of operators which acts on a separable
Hilbert space has a basic measure carried on its spectrum (see
\cite{dix2} Part I, Chapter 7, Propositions 4).

\section{Direct integral of Hilbert spaces}\label{dirint}

General references about the notion of a direct integral of Hilbert
spaces can be found in \cite{dix2} (Part II, Chapters 1-5)  or in  \cite{mau}
(Chapter I, Section 6). In the following we assume that  the pair $\{X,\ssss{B}(X)\}$ is a  standard Borel space and $\mu$ a (regular) Borel measure on $X$. For every
$x\in X$ let $\sss{H}(x)$ be a Hilbert space with scalar product
$(\ ;\ )_x$. The set $\rrr{F}:=\prod_{x\in X}\sss{H}(x)$ (Cartesian
product) is called a \emph{field of Hilbert spaces} over $X$. A \emph{
vector field} $\varphi(\cdot )$ is an element of $\rrr{F}$, namely a
map $X\ni x\mapsto\varphi(x )\in\sss{H}(x)$. A countable family
$\{\xi_j(\cdot )\ :\ j\in\N\}$ of vector fields is called  a \emph{
fundamental family of measurable vector fields} if:
\begin{enumerate}
    \item[{\upshape a)}] for all $i,j\in\N$ the functions $X\ni x\mapsto\left(\xi_i(x);\xi_j(x)\right)_x\in\C$ are measurable;
    \item[{\upshape b)}] for each $x\in X$ the set $\{\xi_j(x )\ :\ j\in\N\}$ spans the space $\sss{H}(x)$.
\end{enumerate}
 The field $\rrr{F}$ has a \emph{measurable structure}
if it has a fundamental family of measurable vector fields. A vector
field $\varphi(\cdot )\in\rrr{F}$ is said to be \emph{measurable} if
all the functions $X\ni
x\mapsto\left(\xi_j(x);\varphi(x)\right)_x\in\C$ are measurable for
all $j\in\N$. The set of all measurable vector fields is a linear
subspace of $\rrr{F}$. By the Gram-Schmidt orthonormalization we can
always build a fundamental family of orthonormal measurable fields
(see \cite{dix2} Part II, Chapter 1, Propositions 1 and 4). Such a
family is called a \emph{measurable field of orthonormal frames}.
Two fields are said to be equivalent if they are equal $\mu$-almost
everywhere on $X$. The \emph{direct integral} $\rrr{H}$ of the
Hilbert spaces $\sss{H}(x)$ (subordinate to the measurable structure
of $\rrr{F}$), is the Hilbert space of the equivalence classes of
measurable vector fields $\varphi(\cdot )\in\rrr{F}$ satisfying
\begin{equation}
\|\varphi(\cdot )\|_\rrr{H}^2:=\int_X\|\varphi(x)\|^2_x\ d\mu(x)<\infty.
\end{equation}
 The scalar product on $\rrr{H}$ is defined by
\begin{equation}
\langle \varphi_1(\cdot );\varphi_2(\cdot )\rangle_\rrr{H}:=\int_X(\varphi_1(x);\varphi_2(x))_x\ d\mu(x)<\infty.
\end{equation}
The Hilbert space $\rrr{H}$ is  often denoted by the symbol $\int^\oplus_X\sss{H}(x)\ d\mu(x)$. It is separable if $X$ is separable.

Let $\nu$ be a positive measure equivalent to $\mu$. The
Radon-Nikodym theorem ensures the existence of a positive $\rho\in
L^1(X,\mu)$ with $\frac{1}{\rho}\in L^1(X,\nu)$ such that
$\nu=\rho\mu$.  Let  $\rrr{H}$ be the direct integral with respect to
$\mu$, $\rrr{K}$ the direct integral with respect to $\nu$ and
$\varphi(\cdot )\in\rrr{H}$. The mapping $\rrr{H}\in \varphi(\cdot
)\mapsto\varphi'(\cdot )\in\rrr{K}$ defined by
$\varphi'(x)=\frac{1}{\sqrt{\rho(x)}}\varphi(x)$ for all $x\in X$ is
an unitary map of $\rrr{H}$ onto $\rrr{K}$ and for fixed $\mu$ and
$\nu$. This isomorphism does not depend on the choice of the
representative for $\rho$ and it is called the \emph{canonical
rescaling isomorphism}.

A \emph{(bounded) operator field} $A(\cdot )$ is a map $X\ni x\mapsto
A(x )\in\bbb{B}(\sss{H}(x))$. It is called measurable if the
function $X\ni x\mapsto\left(\xi_i(x);A(x)\xi_j(x)\right)_x\in\C$ is
measurable for all $i,j\in\N$. A measurable operator field is called
a \emph{decomposable operator} in the Hilbert space $\rrr{H}$. Let
$f\in L^\infty(X)$ (with respect to the measure $\mu$); then the map
$X\ni x\mapsto M_f(x):=f(x)\num{1}_x\in\bbb{B}(\sss{H}(x))$ (with
$\num{1}_x$ the identity in $\sss{H}(x)$) defines a simple example
of decomposable operator called \emph{diagonal operator}. When $f\in
C(X)$, the diagonal operator $M_f(\cdot )$ is called a \emph{
continuously diagonal operator}. Denote by $C(\rrr{H})$ (resp.
$L^\infty(\rrr{H})$) the set of the continuously diagonal operators
(resp. the set of diagonal operators) on $\rrr{H}$. Suppose that
$\sss{H}(x)\neq0$ $\mu$-almost everywhere on $X$, then the following
facts hold true (see \cite{dix2} Part II, Chapter 2, Section 4): (i)
$L^\infty(\rrr{H})$ is a commutative von Neumann algebra and the
mapping $L^\infty(X)\ni f \mapsto M_f(\cdot )\in L^\infty(\rrr{H})$ is a
(canonical) isomorphism of von Neumann algebras; (ii) the commutant
$L^\infty(\rrr{H})'$ is the von Neumann algebra of decomposable
operators on $\rrr{H}$; (iii) the mapping $C(X)\ni f \mapsto M_f(\cdot )\in
C(\rrr{H})$ is a (canonical) homomorphism of $C^\ast$-algebras which
becomes an isomorphism  if the support of $\mu$ is all $X$; in this
case $X$ is the Gel'fand spectrum of $C(\rrr{H})$ and
$\mu$ is a basic measure.

\bibliographystyle{alpha}
\bibliography{biblio_spectral_geometry}
\end{document}